%% file: main.tex
\documentclass[conference]{IEEEtran}
\IEEEoverridecommandlockouts
\usepackage{cite}
\usepackage{amsmath,amssymb,amsfonts}
\usepackage{algorithmic}
\usepackage{graphicx}
\usepackage{textcomp}
\usepackage{xcolor}
\def\BibTeX{{\rm B\kern-.05em{\sc i\kern-.025em b}\kern-.08em
    T\kern-.1667em\lower.7ex\hbox{E}\kern-.125emX}}
 
\usepackage{tikz}
\usepackage{amsmath}
\usepackage{comment}
\usepackage{float}
\restylefloat{table}
\usepackage{float}
\usepackage{textgreek}
\usepackage{filecontents}
 \usepackage{hyperref}

\begin{document}


\title{Evaluating the Effectiveness of Phishing Reports on Twitter }

\author{\IEEEauthorblockN{Sayak Saha Roy}
\IEEEauthorblockA{The University of Texas at Arlington\\
sayak.saharoy@mavs.uta.edu}
\and
\IEEEauthorblockN{Unique Karanjit}
\IEEEauthorblockA{The University of Texas at Arlington\\
unique.karanjit@mavs.uta.edu}
\and
\IEEEauthorblockN{Shirin Nilizadeh}
\IEEEauthorblockA{The University of Texas at Arlington\\
shirin.nilizadeh@uta.edu}
}

\IEEEoverridecommandlockouts
\makeatletter\def\@IEEEpubidpullup{6.5\baselineskip}\makeatother
\IEEEpubid{\parbox{\columnwidth}{
    {\fontsize{7.5}{7.5}\selectfont A camera ready version of this work was presented and will be published at APWG eCrime 2021 – Symposium on Electronic Crime Research  \\
    December 1-3, 2021 \\
    https://apwg.org/ecrime2021\\
  }
}
\hspace{\columnsep}\makebox[\columnwidth]{}}


\maketitle

\begin{abstract} 

Phishing attacks are an increasingly potent web-based threat, with nearly 1.5 million websites created on a monthly basis. In this work, we present the first study towards identifying such attacks through phishing reports shared by users on Twitter. We evaluated over 16.4k such reports posted by 701 Twitter accounts between June to August 2021, which contained 11.1k unique URLs, and analyzed their effectiveness using various quantitative and qualitative measures. Our findings indicate that not only do these users share a high volume of legitimate phishing URLs, but these reports contain more information regarding the phishing websites (which can expedite the process of identifying and removing these threats), when compared to two popular open-source phishing feeds: PhishTank and OpenPhish. We also noticed that the reported websites had very little overlap with the URLs existing in the other feeds, and also remained active for longer periods of time. But despite having these attributes, we found that these reports have very low interaction from other Twitter users, especially from the domains and organizations targeted by the reported URLs. Moreover, nearly 31\% of these URLs were still active even after a week of them being reported, with 27\% of them being detected by very few anti-phishing tools, suggesting that a large majority of these reports remain undiscovered, despite the majority of the follower base of these accounts being security focused users. Thus, this work highlights the effectiveness of the reports, and the benefits of using them as an open source knowledge base for identifying new phishing websites. 
\end{abstract}


\input{introduction}

\input{related-work}

\input{datacollection}

\input{comparison}
\input{interactions}

\input{activity}
\input{conclusion}

\bibliographystyle{IEEEtran}
\bibliography{IEEEabrv,main}

\end{document}

%% file: introduction.tex
\section{Introduction} 

Phishing websites are a prominent social-engineering threat whose volume has significantly increased over the past few years~\cite{phishincreasing:2020,data-breach:2020}. To counter these attacks, there has been a prolonged effort from the security community in the form of automated tools which use sophisticated machine learning~\cite{sahingoz2019machine,rao2019detection,hassanpour2018phishing,abdelhamid2017phishing}, deep-learning~\cite{yang2019phishing,yi2018web,hassanpour2018phishing}, rule-based~\cite{moghimi2016new,sonowal2020phidma,tan2016phishwho,shirazi2018kn0w} and heuristic~\cite{nguyen2013detecting,babagoli2019heuristic,jeeva2016intelligent,sreedharan2016systems} based techniques. But, phishing websites are highly elusive in nature, often evolving to leverage several loopholes and adversarial tactics to circumnavigate automated tool detection~\cite{liang2016cracking,rajivan2018creative,royremains,aleroud2020bypassing}. Thus, domain hosting services, antiphishing tools and web-browsers often rely on one or more phishing feeds/ blocklists, which are curated knowledgebases containing a frequently updated list of phishing URLs, which are either added manually or through a combination of automated discovery/ manual reporting and human verification. As noted by Oest et al.~\cite{oest2020phishtime}, phishing feeds are \textit{reactive} in nature, with a considerable time gap between the appearance of the website and it subsequently being reported. Even then, these feeds usually go through some form of manual evaluation, ideally making them less error prone than automated detection systems. 

In this work, we investigate phishing reports which are shared on Twitter~\cite{twitter:2021}, the popular micro-blogging platform. To the best of our knowledge, this work constitutes the first study on evaluating this resource as a formidable knowledge-base for identifying new phishing websites, and throughout the course of this paper, we concentrate on finding its effectiveness and compare its characteristics and performance to two other popular open source phishing feeds - PhishTank and OpenPhish. More specifically, this paper: (i) Determines the reliability and volume of information shared by these phishing reports, and how they compare against two other open source phishing feeds. (ii) Being hosted on Twitter, these reports can also be visibly interacted upon by other users on the platform, a feature which is not available to the other two phishing feeds that we study. We thus evaluate the frequency of these interactions, including those from domain registrars and organizations which the reported phishing URLs have targeted, and examine the impacts of these interactions on the detection/removal of the reported URLs. (iii) We also analyze the aftermath of sharing these reports, i.e., how long the URLs remain active after getting reported, as well as how quickly anti-phishing tools detect them. Both are factors which can protect the user from inadvertently accessing the threat.

We collected and analyzed more than 16k tweets which contained 11k unique URLs, over the period of 21st June to 17th August 2021. Using a combination of automated and manual investigations, we repeatedly tracked a myriad of properties extracted from these posts including checking the activity of these URLs, anti-phishing tool detection, information shared by these posts (such as relevant hashtags, images, etc.), true positive rate (percentage of URLs which are legitimate phishing attacks), interactions with other users (likes, comments and replies). We also compared the relevant statistics with two other phishing feeds- PhishTank and OpenPhish.

In Section~\ref{datacollection}, we underline how we collected the phishing reports from Twitter, and characterize them based on the domains they are hosted on.
In Section~\ref{driveby}, we discuss about several drive-by downloads URLs which are also reported by these accounts, a feature absent from the other two feeds.
In Section~\ref{comparison}, we evaluate the information shared by these phishing reports (\ref{content-shared-by-reporters}), as well as by PhishTank (\ref{phishtank-share}) and OpenPhish (\ref{openphish-share}). We focus on the information shared by these reports (such as screenshots, IP address, name of targeted registrar/ organization, labelling of threats, etc.), as well as their reliability and validity in Section~\ref{validity-url}. Our sole goal in this section is to evaluate how the information shared by phishing reports on Twitter compare with the offerings from PhishTank and OpenPhish, which is later summarized in \textbf{Table~\ref{functionality}}. Unlike the two other phishing feeds, which usually do not allow user interaction, in Section~\ref{phishing-report-interactions}, we find the volume of interactions (favourites/ retweets/ comments) that these reports get on Twitter, and whether interaction from the targeted domain/organization has an impact on how quickly these reported websites are removed. We also qualitatively explore how these interactions look like (Fig~\ref{fig:interaction}) and determine the technological proficiency of users who typically interact with these reports (\ref{followers}).  Finally, in Section~\ref{activity-website}, we determine how long these URLs stay active after being reported, and how the rate (and pace) of removal compares with URLs which are posted on PhishTank and OpenPhish (Section~\ref{comparison}). We also check for the coverage of the phishing URLs by anti-phishing engines~\ref{virustotal-coverage}. Our main findings can be summarized as below:

\begin{enumerate}
   
\item Twitter is a viable candidate for being utilized as a knowledge-base for phishing reports. Over the course of three months, users consistently shared over 16.4k such reports which covered 11.1k unique URLs hosted over 203 unique registrars, and targeted 146 different organizations. Unlike PhishTank and OpenPhish, these accounts also reported URLs distributing Drive-by downloads (7\%).

\item Majority of the URL reports taken from Twitter contained more information compared to PhishTank and OpenPhish, which can help domain registrars and anti-phishing tools to expedite the process of threat identification.
 
\item The URLs shared in these reports have a high true positive rate (87\%), with only one account contributing to the majority (11\%) of the false positives in our dataset.
 
\item These reports receive very low engagement,  with only 13.8\% of the posts receiving at least one comment. The domain registrars and organizations which the reported URLs targeted (referred to as \textit{targets} henceforth) contributed to only 4\% of these comments. However, when they did respond to these reports, the URLs became inactive more quickly compared to the URLs which do not receive such interaction. Moreover, only 10.2\% of the targets follow at least one such Phishing report account, indicating that they are either not aware of these reporting accounts, or do not consider them as a credible source.

\item 31\% of these reported URLs remain active even after a week of their first appearance in our dataset. Moreover, anti-phishing tools consistently have lower detection rates for at least 27\% of the URLs when compared to URLs which show up in other phishing feeds. 

\end{enumerate}

Thus, our evaluation indicates that phishing reports  shared on Twitter are a reliable and efficient source for conveying information regarding new phishing websites. Relying on these reports can help domain registrars and anti-phishing tools in expediting the process of identifying newer phishing threats. Additionally, based on the volume of information shared by these reports, it proves to be a valuable resource for researchers in building detailed ground truth datasets with less effort and more efficiency compared to other open phishing feeds like PhishTank and OpenPhish. We explore our findings in broader details from the proceeding section on-wards.

%% file: related-work.tex
\section{Background and Related work}

\textbf{Phishing websites:} They are web based threats which usually attempt to trick users into entering their personal information by often pretending to be legitimate organizations. Based on recent measurements, nearly 1.4 million phishing websites are created every month~\cite{14phishing}.  There is no dearth of literature regarding the development of automated phishing detection strategies, including machine learning and deep learning approaches~\cite{sahingoz2019machine,rao2019detection,hassanpour2018phishing}, as well as heuristic and rule-based implementations~\cite{nguyen2013detecting,babagoli2019heuristic,jeeva2016intelligent,mohammad2014intelligent}. However, as highlighted by Vayansky and Kumar~\cite{vayansky2018phishing}, unlike file-based threats, the success of a phishing attack is largely based on human interaction factors, which makes it challenging.  In fact, several qualitative and quantitative studies have determined that users are not proficient at identifying phishing websites~\cite{junger2017priming,bullee2018anatomy,alsharnouby2015phishing,arachchilage2014security,abroshan2021phishing}. Additionally, phishing attacks often evolve based on based on the socio-economic conditions such as the 2008 financial crisis~\cite{phishingfinance:2020}, and more recently, the COVID-19 pandemic~\cite{ahmad2020corona,covidphishincrease:2020}, as well as leveraging several adversarial tactics to circumnavigate automated tool detection~\cite{liang2016cracking,rajivan2018creative,royremains,aleroud2020bypassing}. Thus, domain hosting services, antiphishing tools and web-browsers often rely one or more phishing feeds/blocklists, which are frequently updated knowledge-bases containing a list of new phishing URLs, These URLs are either manually annotated by security conscious individuals or discovered through automated crawlers. Our work focuses on one such knowledge-base distributed across Twitter~\cite{twitter:2021}, the micro-blogging platform, and how it fares against two popular and open phishing feeds - PhishTank~\cite{phishtank} and Openphish~\cite{openphish}.

\textbf{Effectiveness of Phishing feeds:} These are specialize feeds dedicated towards keeping track of new phishing threats which are distributed across the web. These feeds are both closed (proprietary) and open in nature. In this work we only focus on comparing phishing reports posted on Twitter, with two feeds belonging to the open source category - PhishTank (PT) and OpenPhish (OP), since it is easier to collect and analyze a large volume of data from them. Despite the utility of these open phishing feeds, academic research on them is limited, but even then, prevalent work has highlighted several pitfalls that these feeds face consistently. For example, Sheng et al.~\cite{sheng2009empirical} noted how they had a very low efficiency at identifying newer threats at hour zero - The time when phishing threats are at their most potent state and continue to have a low coverage even aftera few hours. Bell et al.~\cite{bell2020analysis} notes that Phishtank and OpenPhish have very few URLs overlapping, suggesting that using them collectively can help in covering a larger volume of these threats. In this work we determine the volume of URLs that are reported exclusively by phishing reports posted on Twitter, and the need for using it as an anti-phishing knowledgebase. Moreover, Moore et al.~\cite{moore2008evaluating} points out the unreliability of PhishTank, one of the most popular community driven phishing feed, because it is prone to false positives and even deliberate poisoning, the former of which we  discuss further in our work. Finally, Oest et al.~\cite{oest2020phishtime} suggests using an \textit{evidence} based phishing reporting feed containing additional artifacts such as screenshots can expedite the process of detecting and removing the threat, which, based on our analysis, neither of the two feeds can do very efficiently at present. Considering that these feeds are being used exhaustively by web browsers~\cite{firefox:2021,operafeed:2020,safarifeed:2020}, as well as anti-phishing tools and organizations~\cite{phishfriends}, these shortcomings can inadvertently impact the protection that is offered to users by these services. By critically evaluating the effectiveness of phishing reports posted on Twitter, we discuss the shortcomings of PhishTank and OpenPhish in both reliability and coverage.

%% file: datacollection.tex
\section{Data collection and characterisation}
\label{datacollection}
To automate the process of collecting tweets which contained phishing reports using the Twitter API~\cite{Twitter:2020}, we qualitatively analyzed 500 random posts containing such reports and identified the  attributes which were unique to them. We found that the majority of such tweets report the URLs in an obfuscated format, usually replacing `http'/`https' with case insensitive variants of `hxxp/hxxps.' This strategy is popularly known as \textit{`URL defanging'}~\cite{urldefanging}, and is used to prevent users from accidentally visiting the malicious link. Thus we utilized two search terms - \textit{`hxxp'} and \textit{`hxxps'} to populate our dataset. However, in some cases, other parts of the URL are also defanged, for example, \textit{http://abc[.].com}, but they were usually accompanied with the hashtags \textit{\#phishing} and \textit{\#scam}. We thus collect tweets using those hashtags as well, and then use a regular expression which reverses the defanging by replacing the obfuscating characters from the URL to make them usable for our experiments. 

We thus utilized this data collection approach to collect new phishing reports every 30 mins from the period of June 21st to August 17th 2021, acquiring 16,486 tweets posted by 701 unique reporters in the process, which contained 11,139 unique URLs. During each 30 min period several companion processes were also run, including tracking whether the URL is active, checking if the URL is present in the phishing feeds provided by PhishTank and Openphish, and also tracking how many anti-phishing engines were detecting the URL by using  VirusTotal~\cite{VirusTotal:2020}. 
VirusTotal~\cite{virustotal} is an online URL scanning tool which scans URLs using 80 different anti-phishing engines, and returns an aggregated total of the engines which detected the URL as malicious. It is used frequently by researchers to create a ground-truth of malicious URLs~\cite{masri2017automated,tanaka2017analysis,sun2016automating,wang2019rmvdroid}.
This approach enabled us to get a full picture of how both registrars and anti-phishing engines reacted to the URLs which were shared by these reports.

Additionally, to evaluate the amount of information shared by the phishing reports, as well as their efficiency (i.e., if the URLs shared were actually phishing websites), we also collected the screenshots of both the tweets as well as the website that they had reported. 
To study how other users reacted towards these reports, we collected a snapshot of all interactions towards these posts at end of every day, which included the comments posted on them, as well as the user ids of the individuals who liked and retweeted them. 
\subsection{Distribution of websites across registrars and targeted organizations}
 
\begin{figure}[h!]
\centering
  \includegraphics[width=0.90\columnwidth]{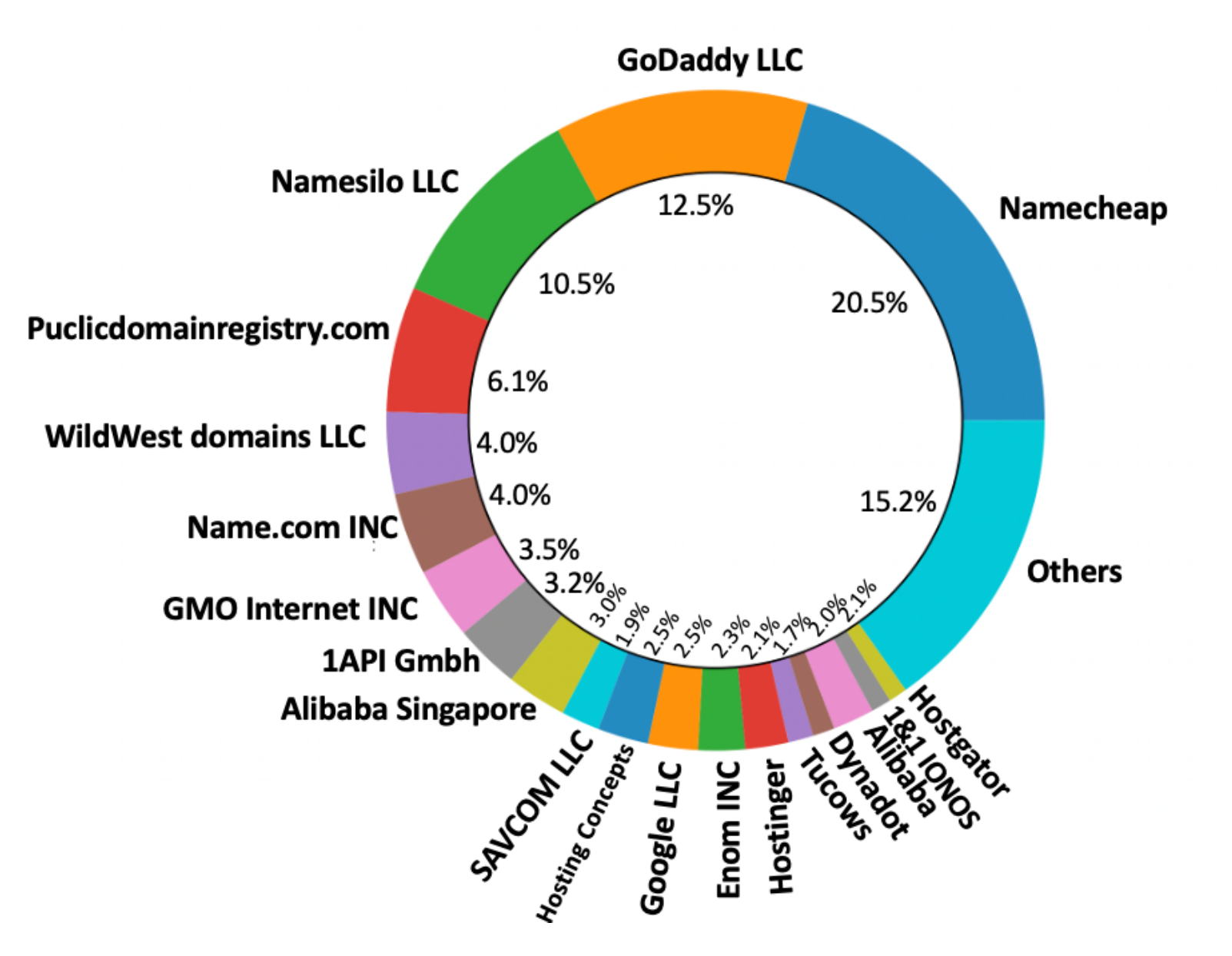} 
\caption{\centering Distribution of URLs across different registrars}
  \label{fig:donutregistrar}

\end{figure}

 We used WHOIS~\cite{whois} to determine the hosting records of the 11,139 unique URLs and found that the reported URLs were distributed across 203 unique domains. Moreover, 5\% (n=631) of the URLs consisted of an adversarial threat category highlighted in work by Saha Roy et al.~\cite{royremains}. These URLs leverage the use of popular free web-hosting domains (which are often white-listed by anti-phishing vendors and phishing feeds alike) to host phishing websites, and in turn remain active for a long time after their first appearance, while also evading detection by several anti-phishing engines. 
 Overall around 52\% of these reports used hashtags (keywords prefixed with \# symbol) to refer to the names of the domains or organization targeted by the URLs. Hashtags are widely used to define a shared context for specific events or topics~\cite{lehmann2012dynamical}, and we assume that the reporters use them to:  (a) inform the domain registrar service and organization that the website is phishing and should be investigated, and (b) inform other users about where the website is being hosted and/or which brand or organization it is targeting, 
 We explore in detail the other informational attributes shared by these reports (such as screenshots of the URL, threat category, location, etc.), and how they compare against PhishTank and Openphish in Section~\ref{comparison}.
 Unlike other phishing feeds, the reports on Twitter can be interacted upon by other users in the Twitter community, including accounts maintained by the domain registrars and organizations which are targeted by the reported URL. Thus, in Section~\ref{phishing-report-interactions}, we explore the responsiveness of these aforementioned parties towards the post, and how it affects the activity of the respective phishing URLs. Figure~\ref{fig:donutregistrar} illustrates the distribution of the URLs across different registrars and drive-by download categories (n=11,139). We find that large amounts of these reported URLs are hosted across popular domain registrars such as GoDaddy, Namecheap, Namesilo, Public domain registry etc. This indicates that these posts are not focused on a particular registrar/ group of registrar, but cover URLs from several sources. Similarly, these reports also cover URLs which host a wide range of file-based threats ranging from Trojan horses, infected PDFs and malicious APKs. We explore the distribution of these threats in Section \ref{driveby}. 

 \begin{figure}[h!]
\centering
  \includegraphics[width=0.80\columnwidth]{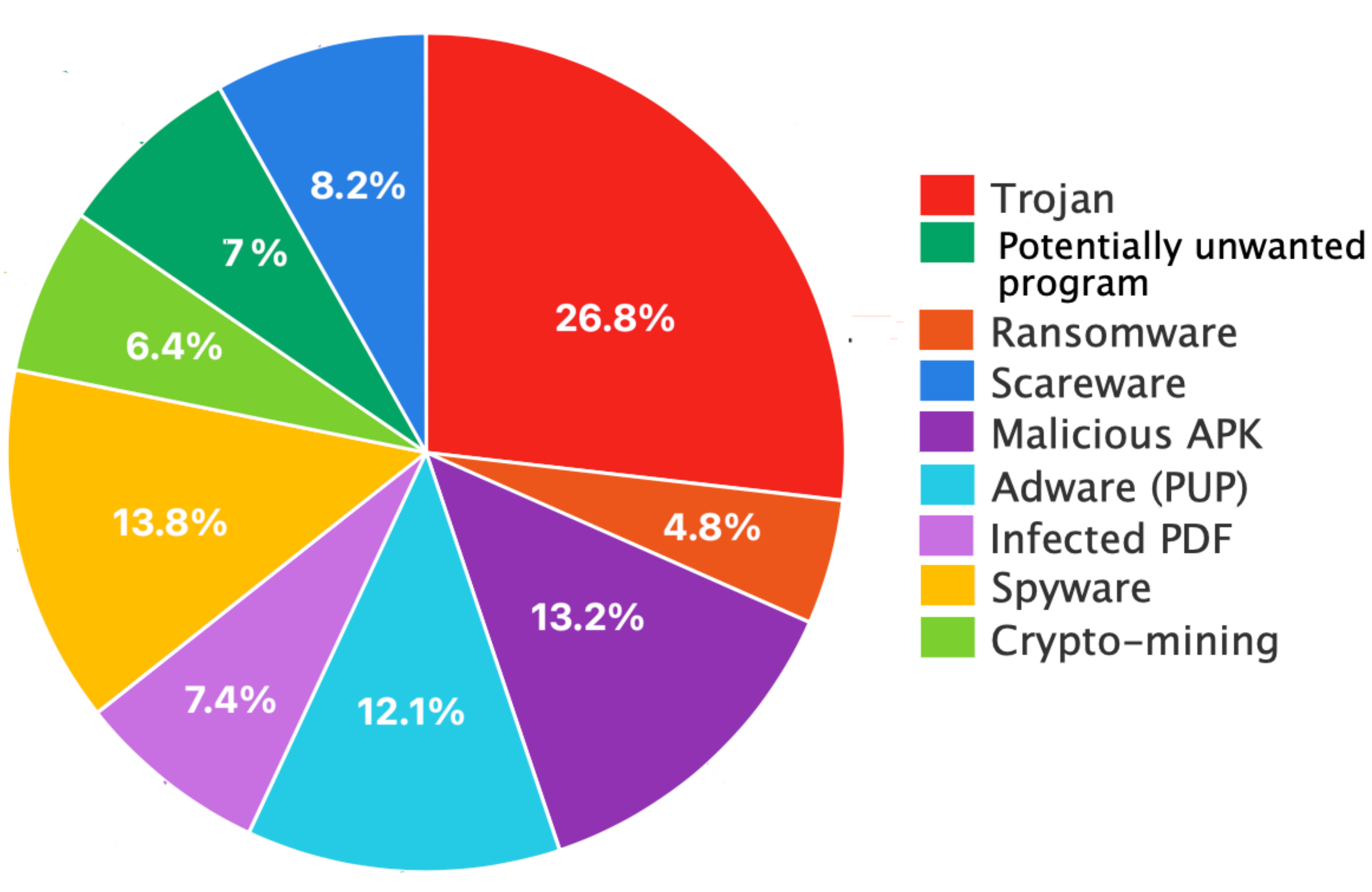}
\caption{\centering Distribution of Drive-by download URLs found in the phishing reports \emph{(n=829)} }
  \label{fig:piedriveby}

\end{figure}

\subsection{Distribution of phishing reporters}
\begin{figure}[h!]
\centering
\includegraphics[width=0.90\columnwidth]{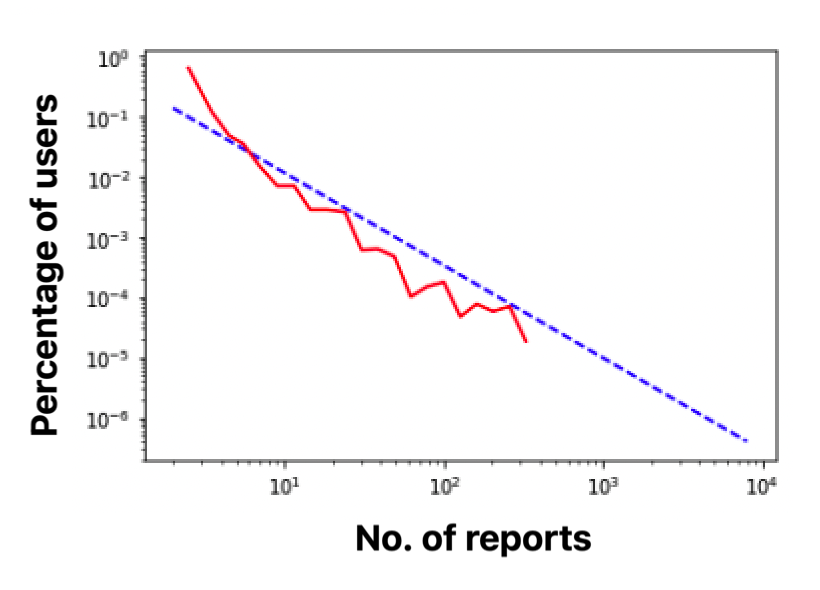} 
\caption{\centering Distribution of the volume of reports shared by the users. }
\label{fig:powerlaw_posts}

\end{figure}

Phishing report tweets in our dataset were posted by 701 accounts. Interestingly, one account posted more than 48.2\% of URLs in our dataset (n=7,946 tweets), 25 accounts posted more than 100 such tweets, and 21 accounts posting more than 50.  Due to only one user contributing such a large portion of the tweets, we report our findings by both considering and not considering this one user (whom we refer to as \textit{top poster} henceforth) separately. Also, 65\% of the users in our dataset shared only one tweet. Infact, the distribution of the posts contributed by these accounts is heavily skewed towards some particular users as illustrated in Figure~\ref{fig:powerlaw_posts}. 
Despite this, our goal is to not concentrate on any one user, but instead investigate the content shared by all these accounts as a form of distributed knowledge-base, and determine the reliability of information provided by these reports and if it can benefit the identification of new phishing threats.

\subsection{Distribution of drive-by download websites}
\label{driveby}
While phishing URLs leverage social engineering techniques using various persuasion principles such as authority and distraction to deceive the users into sharing their private information~\cite{ferreira2015analysis}, websites distributing drive by downloads might contain no phishing component at all, but can still acquire sensitive information by installing malicious files or applications in the user's system and exploiting critical vulnerabilities~\cite{provos2008all}. 
 
We found 829 unique URLs shared over 902 reports which distributed drive-by downloads. We monitored file downloads which were triggered automatically by visiting the URLs in our dataset, and the downloaded files were then scanned using VirusTotal, and were labelled as \textit{Drive-by download} only if those files were detected by at least two different engines (a threshold considered as a standard for labelling malicious files in both the industry and research communities~\cite{peng2019opening}). 
We then distributed these files equally between our team of four security researchers, who executed each of them in a secure VM environment, and based on their characteristics of these files, each file was assigned a label indicative of the threat family they belonged to. We adhered to the threat family labels mentioned in Cisco's Cyber-security Trend report~\cite{ciscotrend}, but also added two more categories which were distinctly present in our dataset: Malicious APK (Android based malware), and Infected PDFs. Figure~\ref{fig:piedriveby} breaks down the distribution of the malicious files across 9 different threat families. 
 We find that 26.8\% of these drive-by downloads distributed trojan horses~\cite{trojanhorses}, malicious files disguised as legitimate software. A good portion (13.2\%) of the URLs distributed Android based malware~\cite{androidmalware}, which ranged from apps which attempted to send premium text messages, showing intrusive advertisements, trying to gain access to system resources etc. We also find cryto-mining malware files, or \textit{crypto-jacking} attacks (6.4\%), which tend to use large amount of system resources to illicitly mine crypto-currency for the attackers gain~\cite{cryptojacking}. 13.8\% of files also consisted of Spywares, ranging from Browser hijackers~\cite{browserhijacker} to Keyloggers~\cite{keylogger}. Also present are Scarewares~\cite{scareware} and Ransomwares~\cite{ransomware} which attempt to which utilize social engineering techniques to restrict/deny control to system data/resource, to threaten the victims into sharing their private information. Thus, it is evident that the Drive-by download URLs shared by the phishing reports on Twitter cover a large array of threat families. 
 In Section~\ref{comparison}, we further look into the coverage of drive-by download URLs by PhishTank and OpenPhish.

%% file: comparison.tex
\section{Comparison with other phishing feeds}
 \label{comparison}
 In this section, we determine the characteristics and volume of information shared by the phishing reports posted on Twitter, and also compare those attributes with the URLs which are shared on PhishTank and OpenPhish in Sections~\ref{phishtank-share},~\ref{openphish-share} respectively. We further extend the comparison in Section~\ref{morecompare} by looking at what portion of URLs overlap between the phishing reports and the two other feeds, as well as what portion of URLs are alive when they are shared. Finally, in ~\ref{validity-url}, we use sophisticated machine learning tools, as well as qualitative analysis to determine what portion of URLs shared by the reporters on Twitter are legitimate(true-positives).
\subsection{Content shared by these reports}
\label{content-shared-by-reporters}
Using regular expressions, as well as extracting the hashtags from these tweets, we were able to analyze the content presented by these reports. Overall, we found that they shared much more information than just the suspected URL. These included the IP address (31\%), hosting registrar(52\%), targeted organization (47\%), the category of the URL (for example phishing, scam or malware - 36\%) as well as a full image (23.5\%) of the phishing website. 
Figure~\ref{fig:reports} provides examples for two such reports and highlights the information shared by them. Without considering the \textit{top poster}, these statistics increased considerably, with 44\% of the posts sharing IP addresses, 61\% and 53\% sharing hosting registrar and domain targets, respectively, 28\% sharing full images of the websites and 42\% sharing the name of the threat. This indicates that the tweets shared by the \textit{top poster} often have less information compared to other reporters. We now consider each of the features (IP address, hosting registrar, targeted organization, etc.) that we have identified from these reports and determine if the other phishing feeds- PhishTank and OpenPhish provide similar information:
  
     \begin{figure}[h!]
\centering
  \includegraphics[width=0.75\columnwidth]{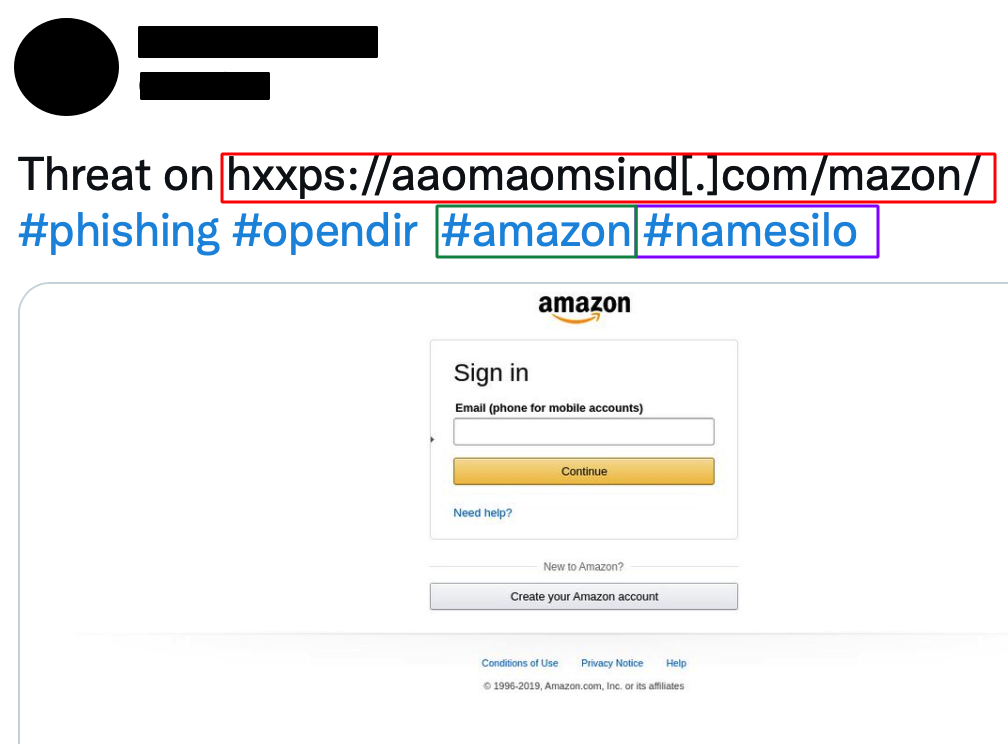}
    \includegraphics[width=0.75\columnwidth]{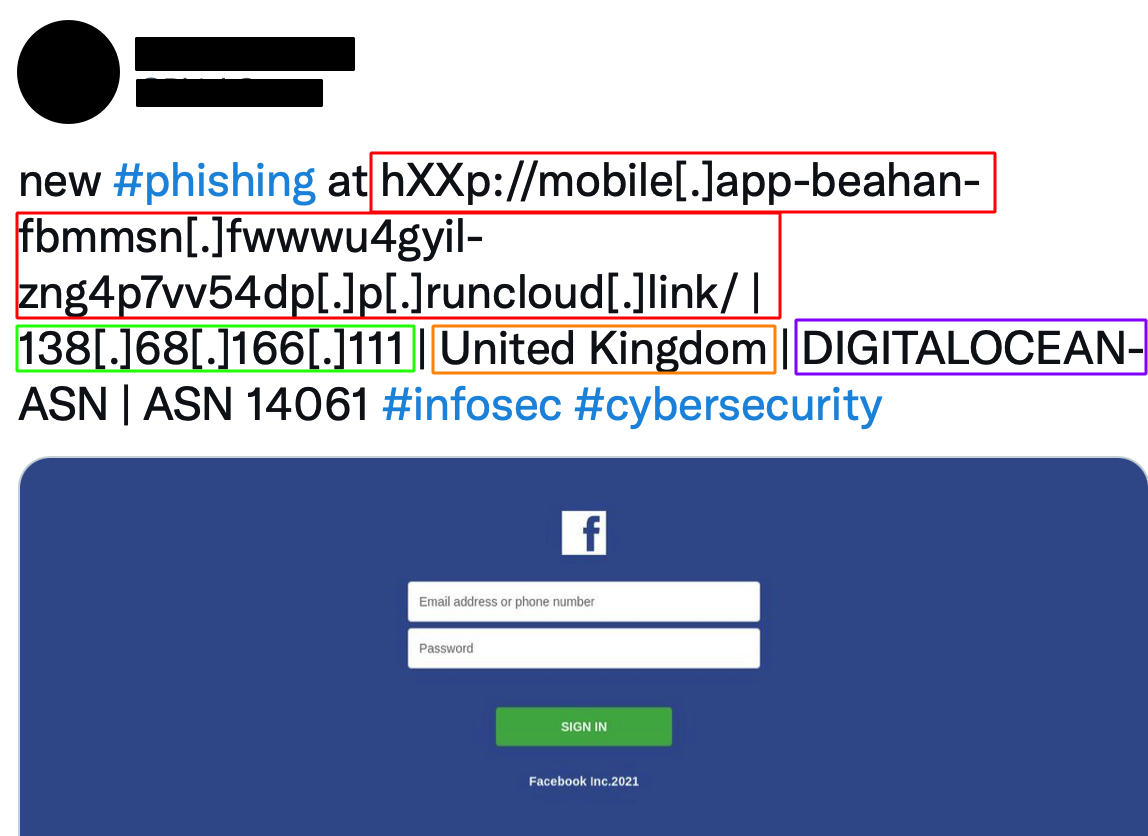}

\caption{Examples of information shared by phishing reports posted by Twitter accounts. \textbf{(On Top)} Report highlights the phishing URL (in red), the targeted organization (in green), and the hosting registrar (in purple). \textbf{On Bottom} In addition to the URL (in red) and registrar hosting (in purple), also shows the Location where the website originated from (in orange), as well as the IP address (in green) } 
  \label{fig:reports}

\end{figure}
 
 \subsubsection{PhishTank}
 \label{phishtank-share}
PhishTank allows any individual to add URLs to their feed, which can then be verified by other users. It does not provide either hosting registrar information nor the IP address of the URLs submitted to their website, and relies only on user submission to populate it's feed. A valid submission only requires the user to provide the URL to be reported and then select the target from a list (with \textit{Other} being a valid option for targets that are not present in the list). They can also provide an open ended response to indicate the contents of the phishing page/email, however this information does not appear anywhere on the feed. Downloading the comprehensive PhishTank feed (which contained 10,622 URLs at the time), we find that nearly 85\% of URLs contained the \textit{Other} label under targeted organization, thus providing no conclusive information about the organization that the URL had targeted.  

Also, since Phishtank relies on  \textit{verifier} accounts to label the submissions on whether the URLs are phishing or not, the feed data provided by PhishTank contains information about when the URL first appeared on PhishTank and when it was verified(the median verification time was around 12.96 minutes). However PhishTank's downloadable feed only provides URLs which have already been verified. Thus, to determine the efficacy of the live feed (which also contains unverified URLs), we first monitored 1k new URLs taken from PhishTank to check what percentage of them are verified. Then through continued observation of these URLs, we found that PhishTank verified nearly 724 of these URLs with a median verification time of 11.49 mins, marking 639 of them as phishing (VALID), and 85 of them as benign (INVALID). However, among the remaining 276 URLs, we found 119 of them to be phishing websites, and we could not observe 37 of them because they were already inactive. Interestingly, among the phishing websites which remained unverified, 53 of them seemed to originate from unconventional phishing domains~\cite{royremains}, a family of phishing threats which are very difficult to detect by both registrars and anti-phishing engines alike. We verified the remaining 120 URLs as false positives, which were added to the 85 URLs that had been labelled by the verifiers as being \textit{INVALID}. Thus, we find that these 1k URLs had a false positive rate of 20.5\%. Considering that researchers often rely on PhishTank as a viable source for collecting phishing URLs~\cite{vtpaper_blackbox}, this rate of false positives might add significant noise to their datasets. 
Also, PhishTank takes screenshots automatically when the URL is submitted (PhishTank does not ask the submitter to provide this information during submission), and if said website is already down then these screenshots do not contribute any useful information towards the appearance of these websites. We found that 29\% of the URLs on PhishTank had screenshots which indicated that the website was already inactive before submission, a phenomenon we investigate more closely in section~\ref{deadonarrival}. Moreover, PhishTank provides a label which indicates whether a URL is \textit{Online}. However, by checking the labels of these 1k URLs, we found that in 33\% of the URLs, PhishTank provides incorrect information about the activeness of the websites (It incorrectly identifies a website is \textbf{Online} when it is actually \textbf{Offline} or vice versa). Thus using this indicator might provide incorrect information about the activeness of the website. 

\subsection{OpenPhish}
\label{openphish-share}
Similar to PhishTank, we focus on 1k URLs which are collected dynamically from the OpenPhish feed. From these URLs, we found that about 39\% of the URLs provided hosting registrar information, and 23\% provided the IP address of the URLs. We also noticed that 74\% of the URLs identified a relevant targeting organization. OpenPhish also reports when a URL first appeared in their feed. To the best of our knowledge, Openphish does not report the screenshot of the webpage, neither on their website, nor through their API access. Unlike Phishtank, Openphish does not report on the activity of the URL as well, i.e., whether a URL is online or not. Also it does not identify the category of the threat of the URL.
How OpenPhish obtains URLs is ambiguous, as they note \textit{``OpenPhish receives millions of unfiltered URLs from a variety of sources on its global partner network.''}~\cite{openphish}. 
However, we assume that these \textit{partners} are curated by OpenPhish themselves, and thus they might be more reliable than the open ended anonymous submission approach implemented by PhishTank. This is further corroborated by the low false positive rate of these submissions, as we identified only 41 (out of 1k URLs) which were incorrectly marked as being phishing. Later on, we sample from this set of URLs in Sections~\ref{activity-website} to track the activity of the reported URLs, as well as how quickly they are detected by anti-phishing engines, and how it compares to phishing reports provided by Twitter accounts.

\begin{table*}[]
\centering
\begin{tabular}{c|c|c|c}
\hline
\textbf{Functionality} & \textbf{\begin{tabular}[c]{@{}c@{}}Twitter Phishing\\ Reports\end{tabular}} & \textbf{PhishTank} & \textbf{OpenPhish} \\ \hline \hline
Submission method & \begin{tabular}[c]{@{}c@{}}Self-submission\\  Self-verification\end{tabular} & \begin{tabular}[c]{@{}c@{}}Community submission\\ Community verification\end{tabular} & \begin{tabular}[c]{@{}c@{}}Partner submission\\ Self verification\end{tabular} \\ \hline 
Hosting Registrar & 52\% \textsuperscript{\textbf{a}}/ 61\%\textsuperscript{\textbf{wt}} & No & 39\% \\ 
Targeted organization & 47\% \textsuperscript{\textbf{a}}/ 53\%\textsuperscript{\textbf{wt}} & 15\% & 74\% \\ 
IP address & 31\% \textsuperscript{\textbf{a}}/ 44\%\textsuperscript{\textbf{wt}} & No & 23\% \\ 
Screenshot shared & 23\% \textsuperscript{\textbf{a}}/ 28\% \textsuperscript{\textbf{wt}}& 71\% & No \\ 
Threat type identified & 36\% \textsuperscript{\textbf{a}}/ 42\%\textsuperscript{\textbf{a}} & No & No \\ 
Drive-by Downloads & 8\% & No & No \\ 
URL Activity status & No & Yes, but error rate of 33\% & No \\ 
Dead on arrival rate & 3.8\% & 24.2\% & 11.4\% \\ 
\begin{tabular}[c]{@{}c@{}}Overlap with Twitter \\ Reports\end{tabular} & N/A & 4\% & 13\% \\ 
False positive rate & 11\% \textsuperscript{\textbf{a}}/ 6\% \textsuperscript{\textbf{a}} & 20.5\% & 4.1\% \\ \hline \hline
\end{tabular}
\caption{Summarizing the information shared by \textbf{Twitter Phishing Reports, PhishTank and OpenPhish} \textbf{a}=Respective stats of all Twitter reports including those from \textit{Top poster}, and \textbf{wt}= Respective stats for all Twitter reports excluding those of top poster }
\label{functionality}
\end{table*}

\subsection{Validity of URLs shared by phishing reporters}
\label{validity-url}
We have established that different phishing feeds share different volumes and variations of information and illustrated how they compare to the Twitter phishing reports. However, since both researchers and industrial entities rely on these feeds up to some capacity, one of the most important aspect of these reports are the validity of the URLs that they share. In the previous section we have already determined that PhishTank and OpenPhish have a false positive rate of 20.5\% and 4.1\% respectively, based on our investigation of 1k URLs collected randomly from these feeds. In this section we evaluate the validity of URLs shared by the phishing reports posted on Twitter. We evaluated the false positive rate of the URLs posted in different report sources  by scanning the URLs on VirusTotal, using manual observation, as well as applying an ensemble machine learning approach. We report the methodology and findings of our evaluation below:
\\ \\
We used VirusTotal as an initial filter to reduce the number of phishing websites needed for manual evaluation. For URLs which had at least 2 detections a day after their appearance in our dataset, we marked them as \textit{true positives}. We found nearly 31\% of the tweets (n=5,109) containing 3,827 unique URLs (34\%) which did not reach this threshold.  Manually labelling such a large volume of URL is not practical, and thus we used two machine learning based implementations, one being a tool developed by Papernot et al.~\cite{npapernot} trained on UCI's Phishing Website Dataset (Mustafa et al.~\cite{mustafadataset}), and the other being Sharkcop~\cite{sharkcop}~\cite{sharkcoppress} to automatically label these URLs. The two different tools were used together for consensus, i.e., a URL was only considered as phishing or benign if both the tools displayed the same label. To gauge the effectiveness of these tools, we manually observed 200 URLs from our dataset and observed an accuracy of 94\% for our setup. Any URL where the tools had disparate labels were put aside for manual labelling. 
In this way our setup was able to mark 2,464 URLs, among which it detected 1,619 URLs as phishing and 845 URLs as benign. The remaining 1,363 URLs were labelled by 4 independent coders. To make sure the coders did not directly interact with the potentially malicious websites, we provided them with screenshots of the website, and the image also contained the URL of the website. The coders verified 824 URLs as phishing and 539 URLs as benign. Thus, for URLs which had less than 2 detections on VirusTotal, we found 2,443 URLs to be phishing and 1,384 URLs to be benign. In total, we found 9,755 URLs to be phishing (87\% of all unique URLs) which were contained in 15,241 tweets.

\begin{figure}[t]
\centering
  \includegraphics[width=0.70\columnwidth]{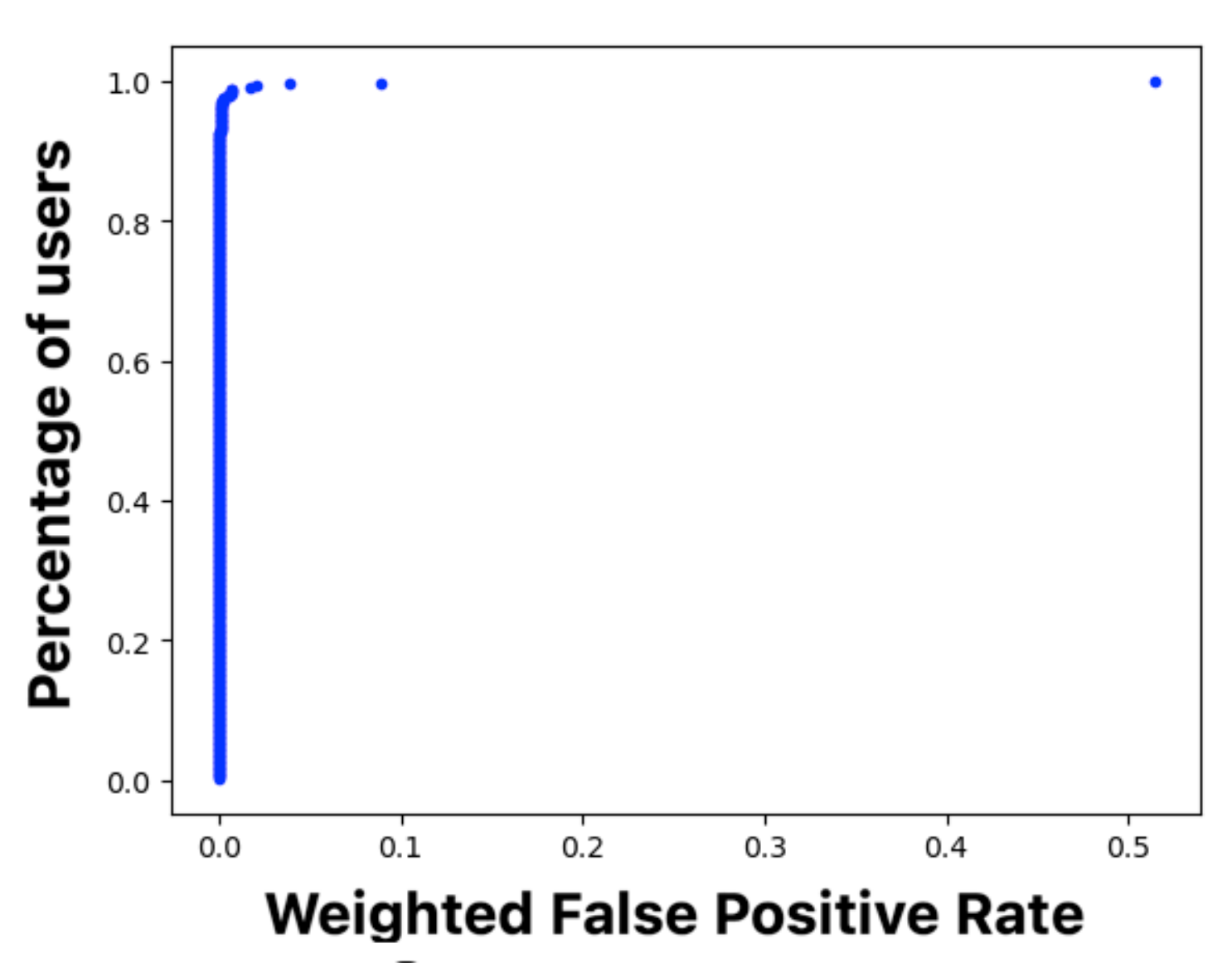} 
\caption{CDF of the weighted false positive rate for each user}

  \label{fig:weightfpr}

\end{figure}

Therefore, it can be established that the URLs reported by these phishing reporters have a high true positive rate. However, our dataset is highly skewed towards the \textit{top user} who contributed 48.2\% of the tweets to our dataset. Interestingly, we found that out of 1,384 benign URLs, 712 URLs were posted by this user alone, which constitutes 11.3\% of all unique URLs posted by this account (n=6,258, out of which 4,188 URLs were unique). As mentioned in Section~\ref{content-shared-by-reporters}, we found that the \textit{top poster} shares fewer details in their reports, compared to other users. Since the distribution of our dataset with respect to tweets shared by the reporters is non-uniform, with a large number of users only sharing one post, we construct a cumulative distribution of the \textit{weighted} false positive rate based on how many posts each user shared versus how many of these shared posts contained URLs which were false positives. We illustrate the distribution in Figure~\ref{fig:weightfpr}. We find that the \textit{top poster} is one of two outliers in the distribution, with only one other user whose feed further contributed to 10\% of the false positives. However, both these users have a high TPR rate of 91\% and 88\%, respectively as well. Outside of these two outliers, as is evident from the diagram, most users have a \textit{false positive} rate of less than 1\%. Thus, the majority of these reporters are much more reliable than PhishTank and OpenPhish with respect to validity of the URLs that they report.

\subsection{Comparison of other attributes}
\label{morecompare}
\subsubsection{Dead on arrival rate (DoA)}
\label{deadonarrival}
We identify a URL as being \textit{dead on arrival} when said URL is already inactive when it first appears on a phishing report/feed. We randomly selected 1k URLs from our phishing report dataset, and checked if they were active when they first appeared in a report, and compared them with 1k URLs we had already selected from  Openphish and Phishtank. We found 24.2\% of those URLs on Phishtank \textit{are dead on arrival}. This statistic is 11.4\% for Openphish. In comparison, only 3.8\% of URLs posted by phishing reporters on Twitter exhibit this behaviour.

This indicates that URLs when posted on Twitter reports are much likely to be alive, and thus need immediate intervention from the targeted registrars and organizations.

\subsubsection{Overlap between reported URLs and other phishing feeds}
\label{overlap}
Considering each URL which were labelled as True Positive in Section ~\ref{validity-url} from the reports in our dataset, we queried their availability on OpenPhish and PhishTank using their respective APIs. Prior literature \cite{bell2020analysis} has noted that URLs keep appearing and disappearing from these phishing feeds, based on if they are still active or not. Thus, we keep checking for the URLs in both OpenPhish and PhishTank every 30 minutes, till after a week of its first appearence in the respective dataset. We find that a low number of URLs overlap with entries on Openphish and PhishTank, with the former having only 13\% of URLs overlapping with the Twitter phishing reports, and the later a mere 4\%. This indicates that a lot of true positive URLs posted by the phishing reporters on Twitter do not appear in either PhishTank or OpenPhish. Interestingly, 5.8\% of the overlapping URLs that showed up in Openphish did so at a median time of 6 hours after being posted on Twitter. The same statistic stands at 1.3\% for Phishtank. While it is difficult to ascertain if these feeds take some input from these phishing reports, our findings do suggest that the phishing reports on Twitter are a faster medium to discover newer phishing threats, as they discover these URLs more quickly than both OpenPhish and PhishTank. Also considering that registrars and even anti-phishing engines often rely (at least partially) on these phishing feeds to identify URLs, these feeds failing to cover a large percentage of tweets found on Twitter can be detrimental for user protection. Thus our findings indicate that the phishing reports are an untapped resource for quickly acquiring a vast breadth of information about new phishing websites when compared to PhishTank and Openphish. We summarize the functionalities exhibited by the reports from each of these phishing feeds in Table~\ref{functionality}. In the next section, we determine how other users on Twitter interact with these phishing reports. 

%% file: interactions.tex
\section{Phishing Report interactions}
\label{phishing-report-interactions}

We collected comments posted on each of the phishing report tweets, and found that only 2,285 tweets got at least  one reply which is around 14\% of the dataset. Moreover, very few of these interactions come from the registrars or the targeted organizations (752 out of 2285 conversations with at least 1 reply, 4\% overall). This is despite the fact that 55.2\% of these reports contain a hashtag citing these concerned services. Figure~\ref{fig:interaction} illustrates the interaction of the registrars/ targeted organizations with the phishing reports. We see that even for services who have more than 100 tweets dedicated to them by the reporters (using hashtags), only 2 of them were able to reply to about 30\% of the tweets that they were tagged in, with 5 targets not replying to any of these tweets. Thus, the CDF indicates that targets have very low interaction with these reports, despite these reports containing URLs which have a high chance of being true positives. We noticed that the median time for getting a reply from the domain registrars is 103 minutes, whereas the same from targeted organizations is is 171 minutes. We term this form of interaction from the registrars/targets as \textit{explicit interaction}, because in these cases, we can say for sure that the target has noticed the report. Later on in Section~\ref{activity-website} we explore how this interaction influences the pace at which these reported websites go offline, and how it compares to posts which do not receive any explicit interaction.

\begin{figure}[t]
    \centering
        \includegraphics[width=0.80\columnwidth]{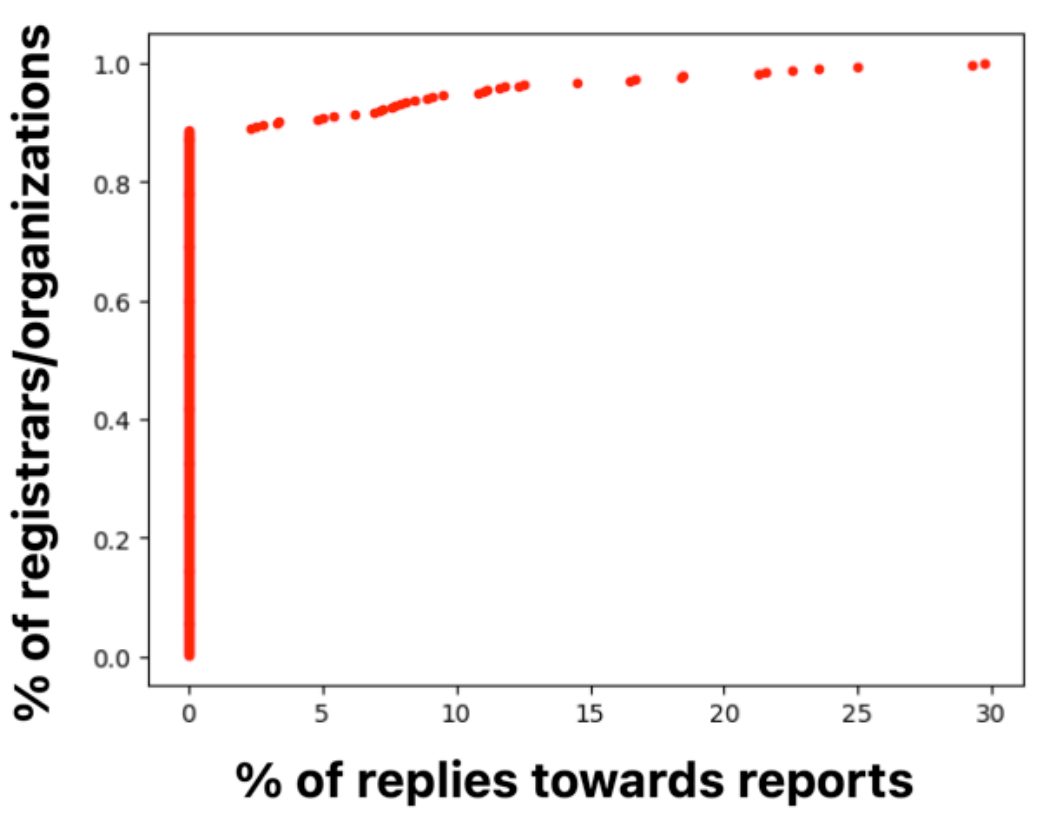}\label{fig:replies1}

        \includegraphics[width=0.80\columnwidth]{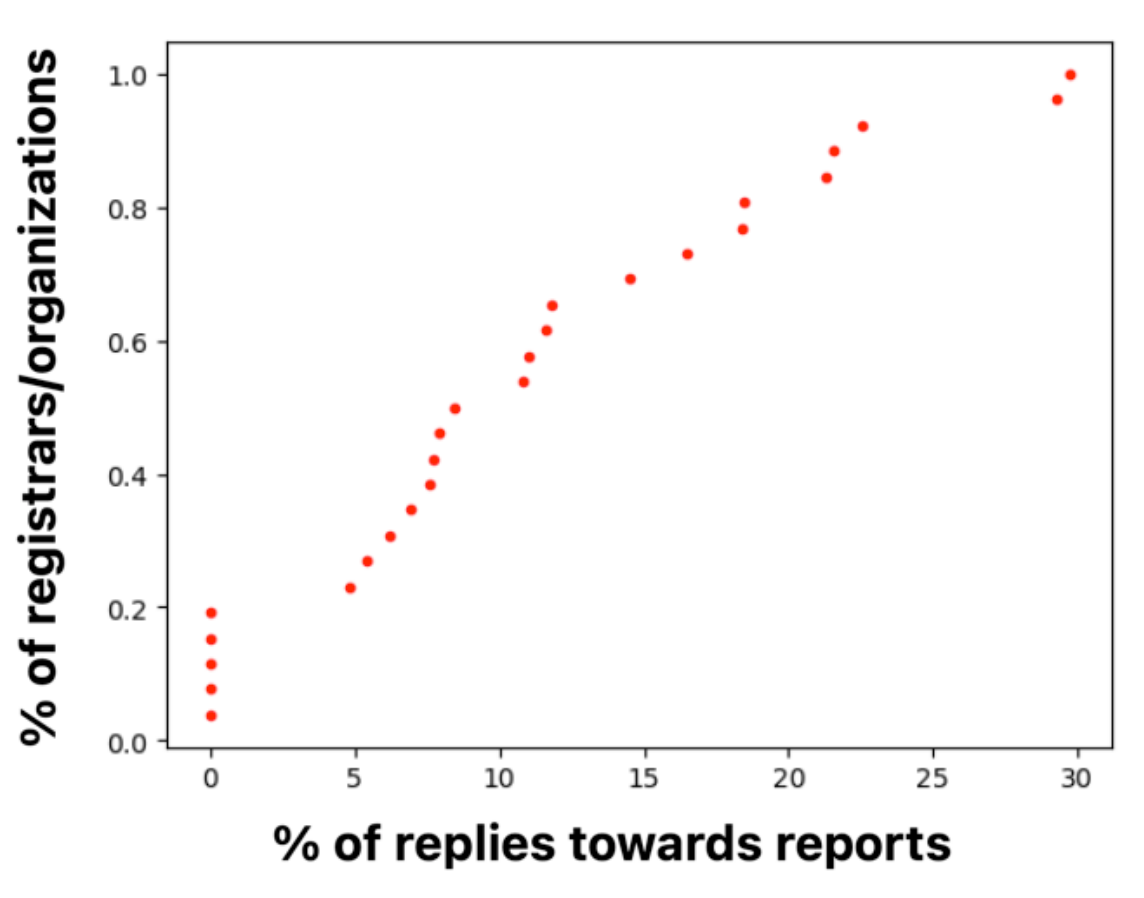}\label{fig:replies2}

         \caption {CDF of replies (explicit interactions) provided by targeted domain registrars/organizations when \textbf{a)} considering all tagged registrars/targets and \textbf{b)} considering only registrars/targets who were tagged in atleast 100 tweets.}
  \label{fig:interaction}
\end{figure}



 \begin{figure}[h!]
\centering
  \includegraphics[width=0.80\columnwidth]{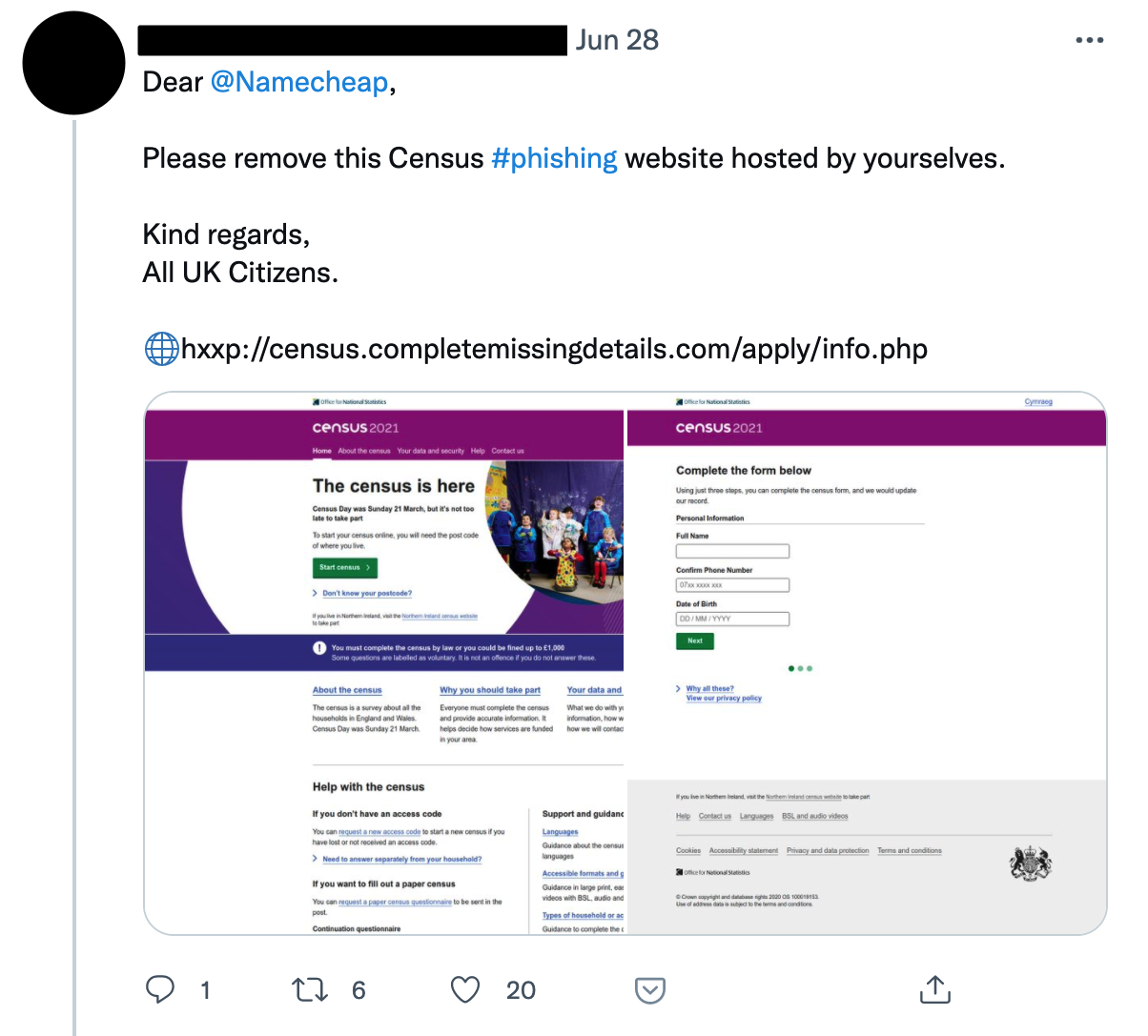} 
  \includegraphics[width=0.80\columnwidth]{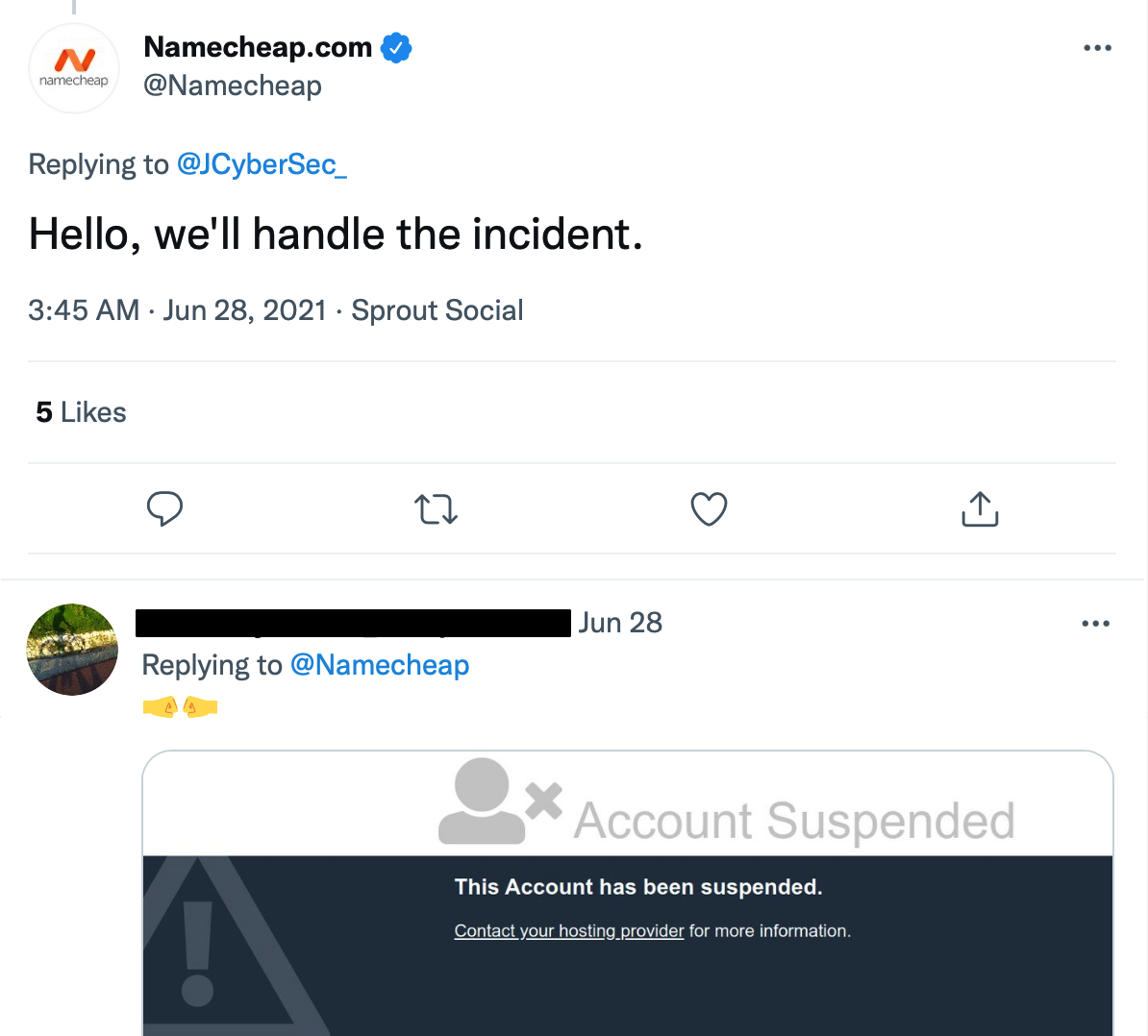} 
\caption{ The effect of explicit interaction from the registrar (On Top) A registrar acknowledges to look into a phishing report. (On Bottom) The reporter confirms that the URL has been removed.}
  \label{fig:interaction}

\end{figure}

\begin{table}[]
\label{descriptive}
\caption{Descriptive statistics of Twitter accounts who shared phishing reports.}
\resizebox{\columnwidth}{!}{%
\begin{tabular}{l|lllll}
\hline \hline
Feature & Type & Min & Max & Mean & Median \\ \hline
Followers & Count & 0 & 127,692 & 1703.78 & 472 \\ 
Posts & Count & 1 & 7,958  & 23.61 & 1  \\ \
Likes & Count & 0 & 205 & 0.45 & 0 \\ 
Retweets & Count & 0 & 161 & 2.07 & 0 \\ 
Listed count & Count & 0 & 6,770 & 83.77 & 7 \\ 
Age (in days) & Count & 42 & 5,298 & 2,325.66 & 2,129    \\ \
Detections & Count & 0 & 23 & 4.09 & 2 \\ 
Verified & Boolean & \multicolumn{4}{c|}{Total accounts = 15} \\ \hline \hline
\end{tabular}}
\end{table}

\subsection{Likes and Retweets}
\label{likesandretweets}
We have already seen in Section~\ref{validity-url} that there is a high chance that the posts shared by the phishing reporting users contain legitimate phishing URLs. Additionally, the URLs posted by these accounts have a low overlap with the URLs posted in other phishing feeds that we have investigated. Thus, the visibility of these tweets is vital to recognize new phishing websites that are reported by them. However we have determined that these reports receive very few interactions from the community, as well as the targeted domains and organizations. Another approach to make sure that these tweets are visible is through likes and retweets~\cite{mcshane2021emoji}. 
 
\begin{figure}[h!]
\centering
  \includegraphics[width=0.40\columnwidth]{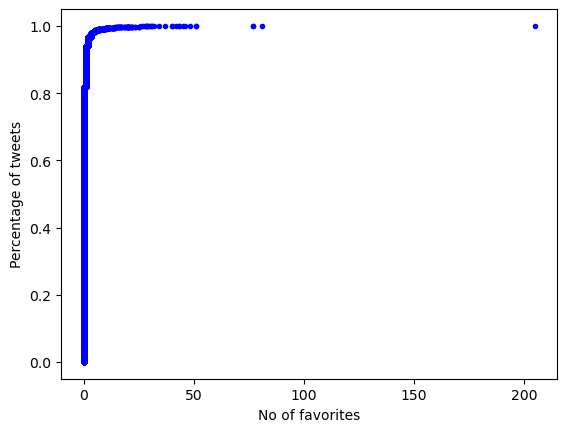}
    \includegraphics[width=0.40\columnwidth]{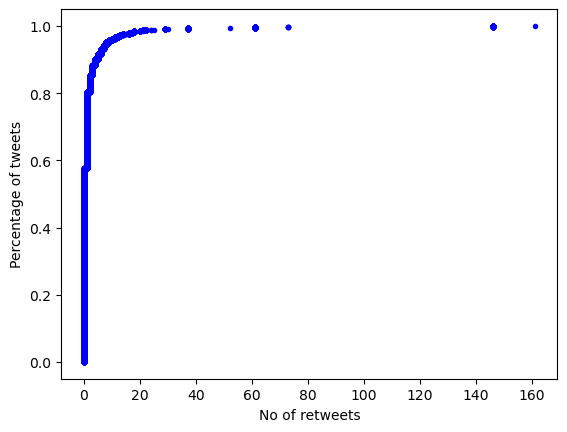}

\caption{CDF of retweets and favourites for Twitter reports }
  \label{fig:ret-fav}

\end{figure}

We found that these posts have very few interactions in the form of likes (median=0) and retweets (median=2) as well. In fact nearly 82\% of the tweets in our dataset (n=13,511) did not receive any likes, and 58\% of tweets in our dataset (n=9,596) did not get retweeted. The cumulative distribution of the number of likes(favourites) and retweets received by the report tweets is illustrated in Figure~\ref{fig:ret-fav}. Thus, the lack of this form of interaction further limits the propagation of these phishing reports through the Twitter community. But, we find that the total number of retweets (n=34,284) is 4.5 times more than the total number of likes received (n=7,527) when all of the tweets in our dataset are considered together. This indicates that the users who interact with these posts have the intention of sharing the information along to their peers, which might lead to more attention towards these tweets. 
  
\begin{figure}[h!]
\centering
  \includegraphics[width=0.80\columnwidth]{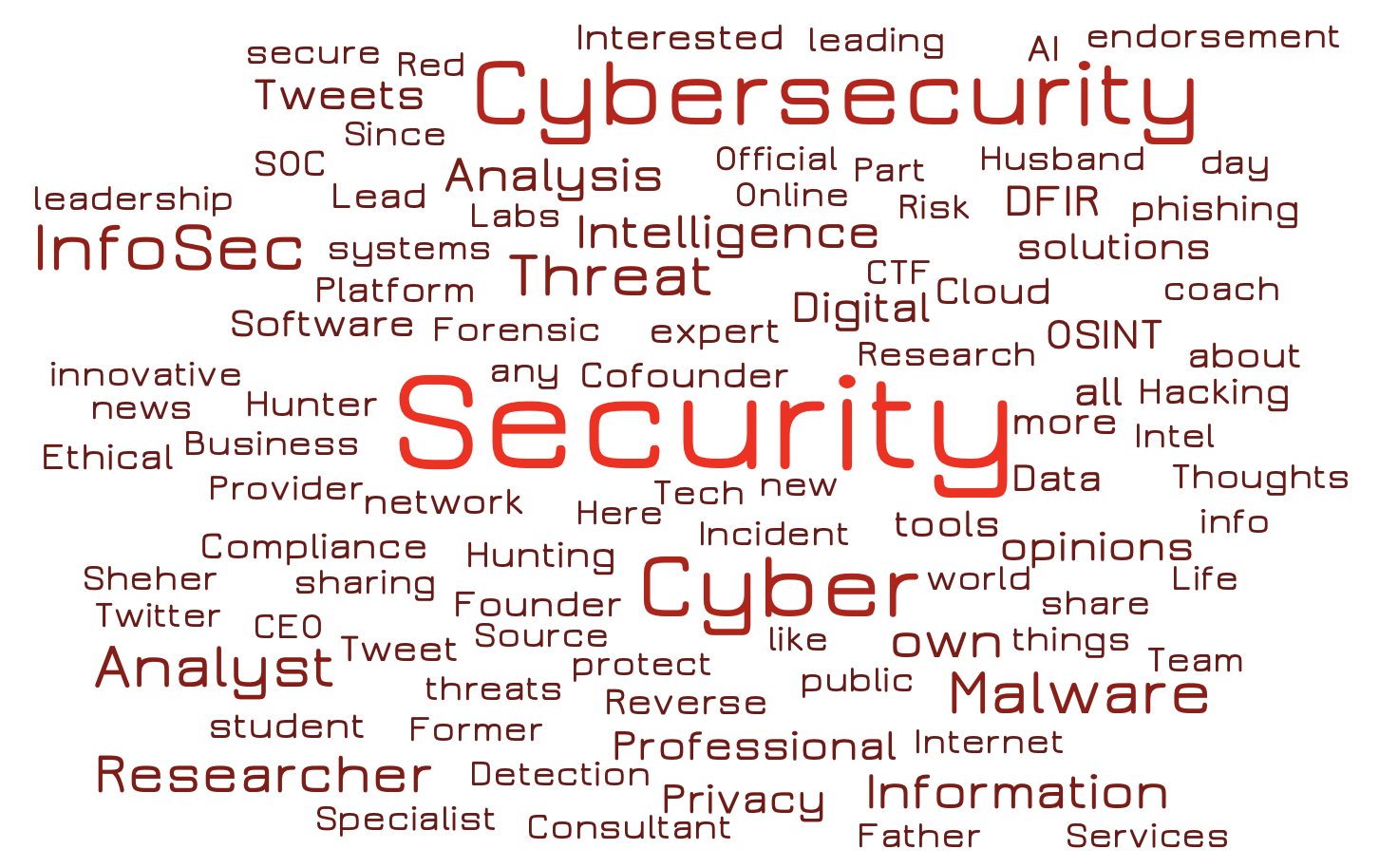} 
\caption{World cloud of the most frequent words found in the profile description of Security focused users who liked/retweeted the phishing report posts}
  \label{fig:wordcloud-security}

\end{figure}

 Now, we are interested in determining what percentage of likes/ retweets received by these tweets are from users in the Technological communities, especially, in the Security field.
 We do so by examining the profile descriptions of the users who have liked and retweeted the tweets in our dataset. Since it is impossible to qualitatively analyze the profile descriptions of all such users, we assigned four coders to go through the profile descriptions of 500 users (who iked/retweeted the reports), to identify which of them indicate that the individual's line of profession/interest is Technological(\textit{Tech}), Computer Security (\textit{Security}) or they are not related to Tech (\textit{Non-Tech}). Based on this labelling, we picked out the profile description of the users marked as \textit{Security} and created a Word-cloud as illustrated in Figure~\ref{fig:wordcloud-security}. 
 We obtain the top 20 most frequently occurring words and their combinations and match it with the profile descriptions of users who liked (n=7,527) and/or retweeted (n=34,284) the phishing reports. We find about 37\% of  likes/ retweets came from users who are interested/work in computer security.  Do note that our findings are based on the keywords that we had selected from the world cloud, and also about 14\% of the users had a blank or irrelevant profile description. Thus, realistically, the number of security focused users who interact with these tweet might be even higher. Even then, a large number of these interactions came from individuals belonging to the security community, which might increase the chances of the reports to be noticed by a registrar/targeted organization. However,  we also note the majority of likes and retweets come from only 5\% of the total no. of accounts that belonged to the security community (based on their profile description).

\subsection{Followers}
\label{followers}

We find that the accounts in our dataset have a median follower count of 472 and median listed count of 7. Despite our previous findings that the phishing reports receive low explicit interaction, as well as very few likes and retweets, the large majority of phishing reporting accounts have a decent number of followers, with 523 accounts having more than 100 followers. On the other hand, we checked the \textit{listed count rate} or LCR, which is the percentage of users who listed an account vs the total number of followers the user has. Listed count is considered as a metric for credibility~\cite{kang2012modeling}, i.e. users tend to list accounts who they rely on for information regarding specific topic(s).
Interestingly, we find that 3 users have a higher LCR than their total no. of followers, however 93\% (n=652) of the reporters have a LCR of less than 10\%, with 40\% of accounts (n=283) having an LCR of less than 1\%. This indicates that despite the users having a decent number of followers, most of them are either not recognized or considered to be a creditable source for providing information, as indicated by their low LCR . Incidentally, the \textit{top poster} account has an LCR rate of only 2.9\% despite contributing the majority of the URLs to our dataset. 
Using the keywords that we had found from the profile descriptions of security related users in Section~\ref{likesandretweets}, we find that at least 33\% of the users belong to the security community. While it is interesting to see that a majority of the users that follow these accounts belong to a relevant community, as we had found earlier in Section~\ref{likesandretweets}, the number of unique users in Security who actually interact with these tweets through likes and retweets is much lower (5\%).

\subsection{Targeted domains/organizations as followers}

\begin{figure}[h!]
\centering
  \includegraphics[width=0.6\columnwidth]{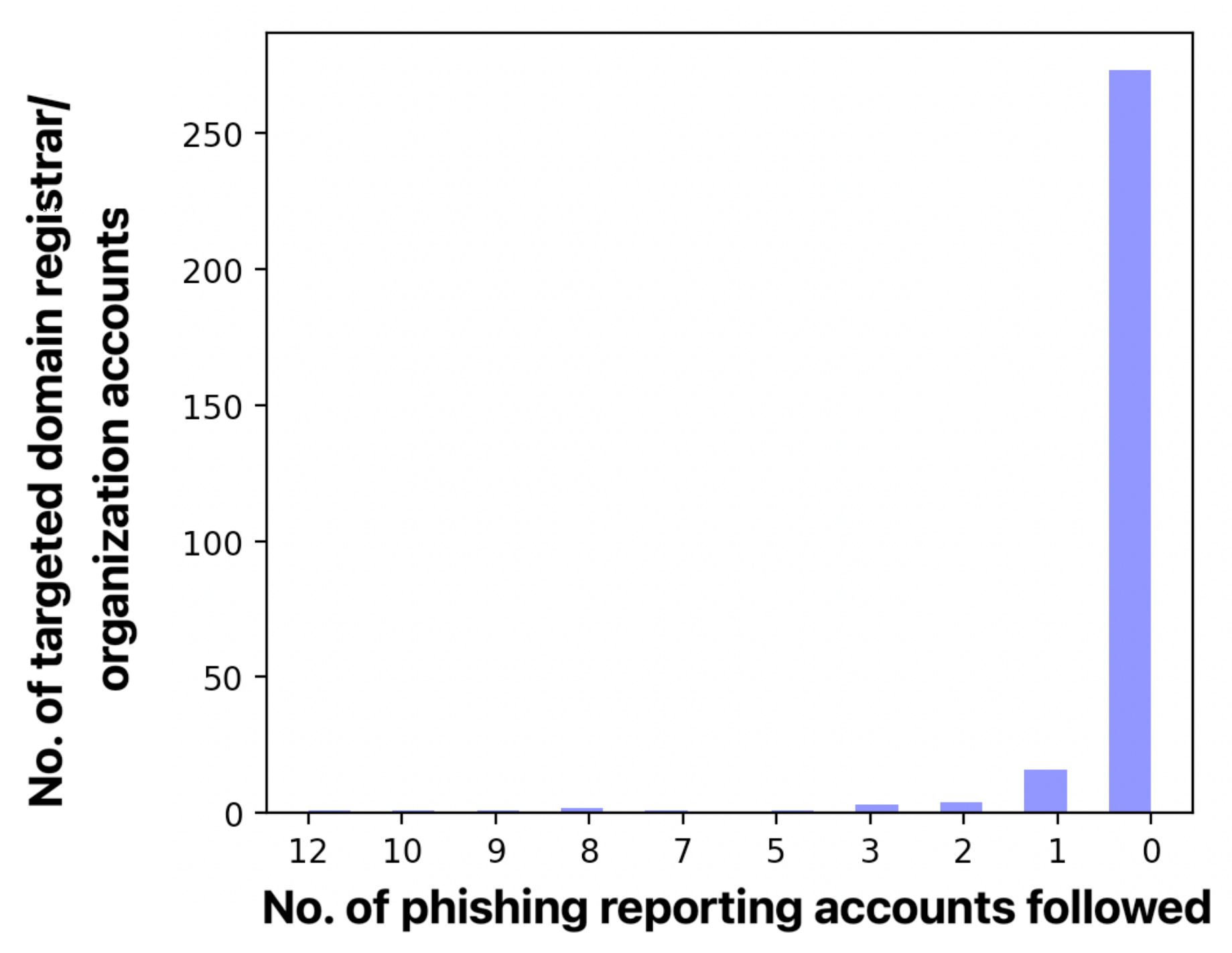} 

\caption{Distribution of the domains/organization accounts across the number of phishing report accounts that they follow.}
  \label{fig:followers-histogram}
\end{figure}

We have already observed that the domain registrars and organizations which are targeted by the reported URLs have very low \textit{explicit interactions} (posting comments) with said reports. But since we have already established that these reports are reliable and provide a lot of information about the phishing website, it is very important that these reports are \textit{discoverable}, i.e., the targeted entities can notice these reports such that they can expedite the process of removing the URLs. The most convenient way to discover such new reports is to follow the phishing report accounts, as posts from these accounts will then show up in the personalized feed of the followers. Out of the 349 registrars and organizations that were tagged by these reports, 303 of them (87\%) have an account on Twitter. Using the Twitter API, we collected the names of all followers for each of our 701 accounts, and then looked for the presence of these 303 registrars/ organizations' twitter accounts in their follower list. We found only 31 targets/ organizations (10.2\%) which follow at least one of the phishing reporting accounts, with only one user following a maximum of 12 accounts. Figure~\ref{fig:followers-histogram} illustrates the distribution of the domains/ organizations across the number of phishing report accounts that they follow. While it is difficult to ascertain how and whether registrars/ organizations keep a track of URLs shared by these phishing reports, our findings imply that a large majority of targeted domains/ organizations do not follow these reporting accounts, either because they are not aware of them, or do not consider them a creditable source for obtaining phishing reports.

\subsection{Age of accounts}

The age of an account denotes how long it has been active on Twitter. We found that the phishing reporting accounts in our dataset have been active for a median period of 2,129 days (5.83 years), with only 83 accounts (12\%) having an age of less than a year. Prior literature has recognized accounts which tend to distribute spam and misinformation to have low account age ~\cite{gupta2018framework,herzallah2018feature}, and thus the longevity of these accounts can be used as yet another feature/indicator by domain registrars and anti-phishing tools to determine whether they should rely on the reports.

%% file: activity.tex
\section{Activity of URLs in phishing report}
\label{activity-website}

We continuously checked whether each unique URL reported in the phishing reports was active and found that throughout the duration of the study, 39\% of URLs reported by the accounts were still active after a day, and 31\% after a week. Since only 752 tweets received a reply from the registrar/ targeted organization (4\% of all tweets in our dataset, containing 671 unique URLs), we compare these tweets with the same number of randomly selected unique URLs included in reports which did not receive a reply from the registrar/ targeted organization. 
Note that for the latter group, we only selected URLs which had become inactive. We performed a Mann-Whitney U Test~\cite{mcknight2010mann} on both groups of URLs, and found that URLs found in reports which get a reply from the targeted organization are more likely to become inactive sooner than URLs in reports which do not get a reply (p$<$0.01). Statistically, URLs in posts  which got a reply from an organization all became inactive within a median time period of 403 minutes. On the other hand, for URLs which did not get a reply, we found the median time of removal to be at 1,172 minutes. However, the latter group of URLs can also be bifurcated into two more groups. Earlier we have seen that 52\% of these reports use a hashtag which cites the registrar or targeted organization. Thus to determine if there is a difference between the activity time of URLs which contained relevant hashtags versus those which did not, we randomly selected 500 posts (each containing unique URLs) from the two groups, and performed a Mann-Whitney test again. Our results indicate that posts which tag the hosting or targeted organization are more likely to be removed more quickly than posts which do not contain such hashtags (p$<$0.01). The median time of removal for URLs which contained a relevant hashtag was at 847 minutes, while those which did not had a median time of removal of 1,591 minutes. It is to be noted that all URLs which garnered a reply from the target had relevant hashtags. However these reports were only 8\% of the overall tweets which had used hashtags (752 out of 8,572 posts), indicating that the majority of reports with hashtags do not recieve an explicit interaction from the target or hosting organizations. Thus, it is hard to determine what factor determines whether a targeted registrar/organization will reply to these reports. 

\subsection{Removal rate comparisons with other phishing feeds}
\label{removalrate}
Phishing report tweets become inactive the quickest when they receive a reply from the targeted registrar/ organization, with the 671 unique URLs removed at a median time of 403 minutes. Comparing this time with the same number of (true positive) URLs chosen from PhishTank and OpenPhish, and also performing respective Mann-Whitney U Tests between removal times between them and the phishing reports, we notice that URLs found on PhishTank are removed at a median time of 132 minutes, having a significant edge over Twitter URLs (p$<$0.05). The same is seen for OpenPhish URLs are removed at a median time of 71 minutes (p$<$0.01). Our findings thus suggest that URLs appearing in these feeds get removed much faster than those found in the phishing reports. However, in Section~\ref{deadonarrival}, we have already found that several URLs submitted on Phishing and OpenPhish are dead on arrival, when compared to those found on Twitter. This when added to the fact that there is minimal overlap of URLs included in the Twitter Reports with the two other phishing feeds, further suggests that: (a) Phishing reports on Twitter are a viable solution for finding new phishing URLs, which are mostly not found on atleast two other popular phishing feeds, and (b) Registrars and targeted organizations are slower at removing websites which show up on these reports, something that can be easily improved upon. We expand upon these findings in Section~\ref{virustotal-coverage}, where we compare the coverage of URLs in phishing reports by anti-phishing tools, compared to those found on the two other phishing feeds.

\subsection{URLs which remained active after a week}

Around 31\% of unique URLs (n=3,453 URLs) remained active even after a week. It is interesting to note that none of the 671 URLs which were part of the posts that targets replied to are found in this category, suggesting that an \textit{explicit interaction} from the targeted registrar/ organization leads to the removal of the website. Almost 67\% of the URLs which did not get removed after a week (n=2,311 URLs) were those which did not have any relevant hashtags. Thus, we hypothesize that the lack of such hashtags might make it difficult for targets to search for them/index them, compared to those which already have a hashtag. The higher rate of removal for URLs which were part of reports which contains relevant hashtags further hints at the phenomenon of \textit{Implicit} interactions between the target registrars/ organizations with these phishing reports, that the latter can investigate (and remove these URLs) without directly interacting with the reports. However, this assumption is not comprehensive, as URLs in posts which had hashtags might have also shown up in phishing feeds which we have not covered in this study, something we can clarify in a future study by focusing only on URLs which exclusively appear in these phishing reports. 

\subsubsection{The case of unconventional phishing URLs}

Work by Saha Roy et al.~\cite{royremains} explored a new category of phishing URLs which use free hosting domains to remain undetected from anti-phishing tools, and are similarly not removed by registrars for a long period of time, if at all. In our phishing report dataset, we found 5\% (n=631) URLs belonging to this category, out of which 53 received a reply from the registrar/ targeted organization. We found that all of these URLs were removed at a median time of 319 minutes, which is much quicker (by several days) than what has been previously established. We selected 100 random URLs  from each of the other phishing feeds which were part of this category, and noticed that such URLs on PhishTank are removed after a median time of 1047 minutes after appearing, whereas the statistics for OpenPhish is 892 minutes. Thus, our preliminary analysis suggests that phishing reports are a much quicker way of making sure these hard to detect URLs are removed, when compared to OpenPhish and PhishTank.

\subsection{VirusTotal coverage}
\label{virustotal-coverage}
In this section, we investigate how quickly anti-phishing URLs pick up on URLs shared by phishing reports and how it compares to OpenPhish and PhishTank.

\begin{figure}[t]
\centering
  \includegraphics[width=1.0\columnwidth]{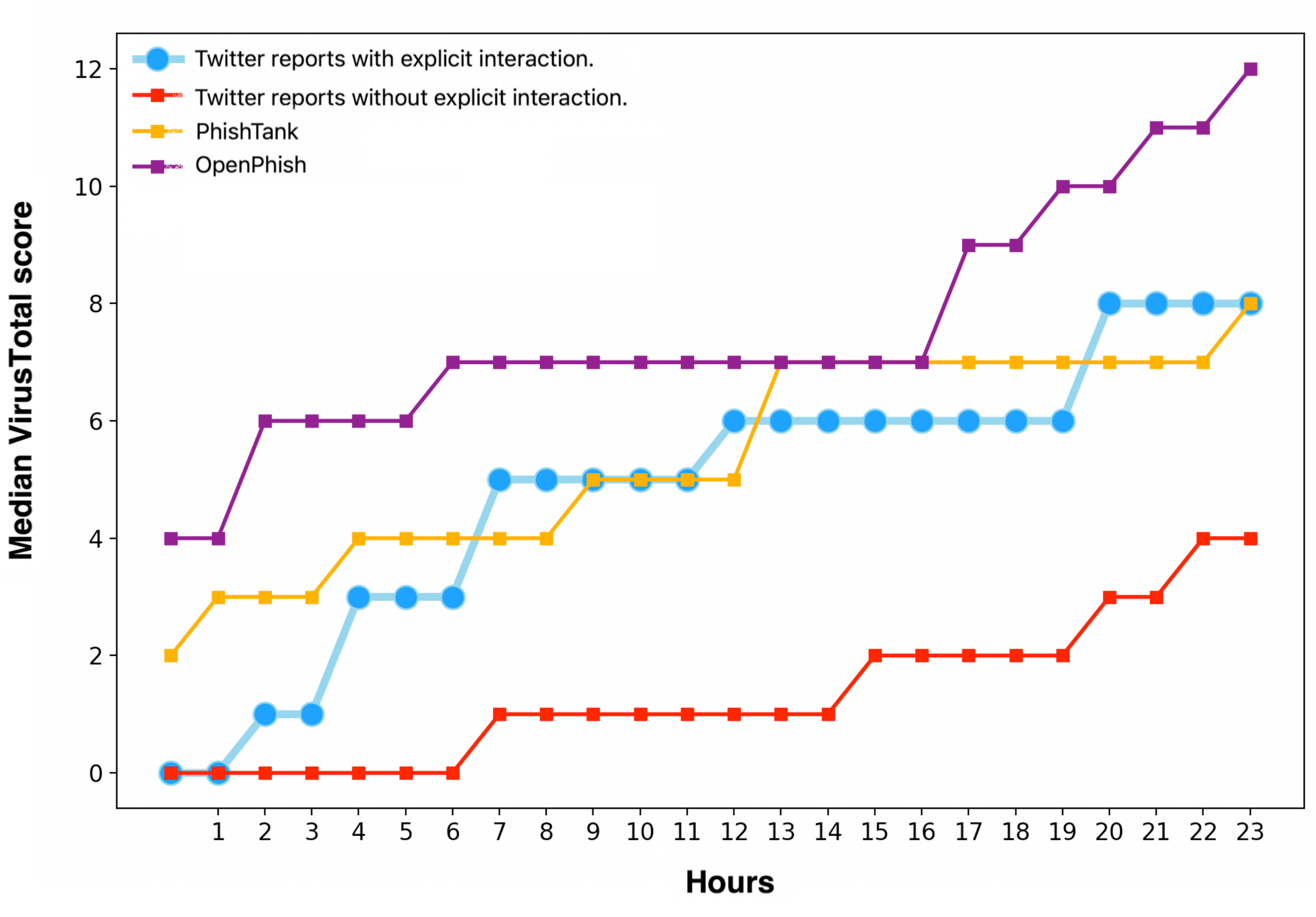}

\caption{Tracking median VirusTotal scores for the reported URLs through their first day of appearance for phishing reports which received a comment from the registrar/ target organization, reports which do not, as well as PhishTank and Openphish.}
  \label{fig:vtotalmedian}

\end{figure}

Since we only have 752 posts on Twitter containing 671 unique URLs which received \textit{explicit interaction} from the \textit{targets}, it would not be fair to compare them with a large volume of URLs from the other conditions, i.e., Twitter posts without explicit interaction posts. Thus, we sample 500 random tweets from each of these sets. Since most phishing URLs and campaigns are only online for less than a day, we tracked how many anti-phishing tools detected these URLs at an interval of every 30 mins through a period of 24 hours. We illustrate the detection of the URLs through time in Figure~\ref{fig:vtotalmedian}. To avoid congestion in the figure we extended the time bins to 1 hr instead of 30 mins.

Our results indicate that URLs on Openphish and PhishTank are detected by a lot more engines within a short time after their appearance compared to those included in Twitter phishing reports. However, we see that reports once explicitly interacted upon by targeted registrars and organizations, see a rapid rise in detection rate by anti-phishing engines, going almost head to head with the other phishing feeds, if not exceeding them. However, URLs which did not get explicit interactions tend to consistently have lower anti-phishing tool detection throughout the day. We have already noticed in Section~\ref{activity-website} that the majority of the URLs included in Twitter phishing reports were alive or had a very slow rate of removal. We see here that they are very sparsely detected by anti-phishing tools as well.


%% file: conclusion.tex
\section{Limitations}

To reliably analyze the qualitative content shared by the tweets used in this study, we only examined reports which were in English. We also limit our data collection to tweets which contained four terms/ hashtags - 'hxxp', 'hxxps', \#phishing and \#scam. Thus, our dataset is not exhaustive, with the possibility of more phishing reports existing in other languages or having different format of text. We also do not account for how the phishing reporters on Twitter obtain the URLs in the first place. Thus as a future study, we propose a qualitative line of research which can investigate these reporters on a case by case basis by conducting interviews and questionnaires to understand how they maintain their accounts on Twitter. Outside of Twitter, we only examine two other open phishing feeds, and since we did not attempt to study proprietary (closed) phishing feeds, it is not possible to determine the contribution of the phishing reports which did not receive any interaction from the targeted entities but got remove anyway, as said URLs may have showed up in those closed feeds, which might have triggered their removal. 

\section{Conclusion and Discussion}

In this study, we establish phishing reports posted on Twitter as a new and reliable resource for sharing information regarding these web-based threats. When compared to two other open phishing feeds - PhishTank and OpenPhish, our findings indicate that phishing reports on Twitter tend to share more information about the phishing URLs, cover an extra threat category (drive-by downloads) of attacks, and tend to have lower volume of false positives. Considering that the best defense against phishing websites is taking them offline, several of these reports also use hashtags to notify the domain registrars and targeted organizations that they have been targeted by the reported URLs. When the targeted entities (\textit{targets}) interact (comment) with the posts, it leads to quick deactivation of the reported URL, as well as getting detected by more anti-phishing engines. However, these interactions were noticed in only 4\% of the posts in our dataset. Among the URLs contained in reports which did not get any interactions, nearly 31\% of them were still active even after a week, in addition to being detected by fewer anti-phishing engines compared to URLs posted on OpenPhish and PhishTank. We also noticed that only 10.2\% of the targets follow at least one phishing reporting account, with only one entity follow a maximum of 12 reporting accounts. This indicates that, despite the majority of these tweets reporting true positive phishing threats, targeted entities are either not aware or do not find these reporting accounts as a viable source for gathering information about new phishing URLs. Additionally, the majority of users, who follow these accounts, belong to the security community, yet a very small minority(~5\%) actually interact or share these reports, which might negatively impact the discover-ability of these reports.

Thus, our evaluation brings to light the effectiveness of phishing reports that are hared on Twitter. The reliability and volume of information shared by these reports can expedite the process of moderation and removal of newer phishing threats. Also, considering the low rate of false positives, security researchers can especially benefit from extracting information from these reports and utilize it for ground-truth labelling. Prevalent anti-phishing tools can also look to extract information from this resource to enhance their own blocklists. However, the current situation indicates that domain registrars and organizations targeted by phishing threats tend to ignore or are not aware of these reports, despite several of the reported URLs being exclusive to these phishing reports and do not show up on PhishTank or OpenPhish. This is further exacerbated by the fact that security focused users who follow these accounts tend not to share these tweets through their network to raise awareness and discover-ability of the reports. Thus, we hope our findings in this work can raise awareness towards the effectiveness of this existing knowledge-base on Twitter, such that it can be integrated in prevalent phishing moderation and research workflows, as well as motivate further research towards analyzing these accounts, and if similar useful knowledge-bases can be encountered within other Online social media networks.

%% file: main.bbl
\begin{thebibliography}{10}
\providecommand{\url}[1]{#1}
\csname url@samestyle\endcsname
\providecommand{\newblock}{\relax}
\providecommand{\bibinfo}[2]{#2}
\providecommand{\BIBentrySTDinterwordspacing}{\spaceskip=0pt\relax}
\providecommand{\BIBentryALTinterwordstretchfactor}{4}
\providecommand{\BIBentryALTinterwordspacing}{\spaceskip=\fontdimen2\font plus
\BIBentryALTinterwordstretchfactor\fontdimen3\font minus
  \fontdimen4\font\relax}
\providecommand{\BIBforeignlanguage}[2]{{%
\expandafter\ifx\csname l@#1\endcsname\relax
\typeout{** WARNING: IEEEtran.bst: No hyphenation pattern has been}%
\typeout{** loaded for the language `#1'. Using the pattern for}%
\typeout{** the default language instead.}%
\else
\language=\csname l@#1\endcsname
\fi
#2}}
\providecommand{\BIBdecl}{\relax}
\BIBdecl

\bibitem{phishincreasing:2020}
F.~Simon~Chandler, ``Google registers record two million phishing websites in
  2020,''
  \url{https://www.forbes.com/sites/simonchandler/2020/11/25/google-registers-record-two-million-phishing-websites-in-2020/?sh=23b004881662}.

\bibitem{data-breach:2020}
M.~Rosenthal, ``Must-know phishing statistics: Updated 2020,''
  \url{https://www.tessian.com/blog/phishing-statistics-2020/}.

\bibitem{sahingoz2019machine}
O.~K. Sahingoz, E.~Buber, O.~Demir, and B.~Diri, ``Machine learning based
  phishing detection from urls,'' \emph{Expert Systems with Applications}, vol.
  117, pp. 345--357, 2019.

\bibitem{rao2019detection}
R.~S. Rao and A.~R. Pais, ``Detection of phishing websites using an efficient
  feature-based machine learning framework,'' \emph{Neural Computing and
  Applications}, vol.~31, no.~8, pp. 3851--3873, 2019.

\bibitem{hassanpour2018phishing}
R.~Hassanpour, E.~Dogdu, R.~Choupani, O.~Goker, and N.~Nazli, ``Phishing e-mail
  detection by using deep learning algorithms,'' in \emph{Proceedings of the
  ACMSE 2018 Conference}, 2018, pp. 1--1.

\bibitem{abdelhamid2017phishing}
N.~Abdelhamid, F.~Thabtah, and H.~Abdel-jaber, ``Phishing detection: A recent
  intelligent machine learning comparison based on models content and
  features,'' in \emph{2017 IEEE international conference on intelligence and
  security informatics (ISI)}.\hskip 1em plus 0.5em minus 0.4em\relax IEEE,
  2017, pp. 72--77.

\bibitem{yang2019phishing}
P.~Yang, G.~Zhao, and P.~Zeng, ``Phishing website detection based on
  multidimensional features driven by deep learning,'' \emph{IEEE Access},
  vol.~7, pp. 15\,196--15\,209, 2019.

\bibitem{yi2018web}
P.~Yi, Y.~Guan, F.~Zou, Y.~Yao, W.~Wang, and T.~Zhu, ``Web phishing detection
  using a deep learning framework,'' \emph{Wireless Communications and Mobile
  Computing}, vol. 2018, 2018.

\bibitem{moghimi2016new}
M.~Moghimi and A.~Y. Varjani, ``New rule-based phishing detection method,''
  \emph{Expert systems with applications}, vol.~53, pp. 231--242, 2016.

\bibitem{sonowal2020phidma}
G.~Sonowal and K.~Kuppusamy, ``Phidma--a phishing detection model with
  multi-filter approach,'' \emph{Journal of King Saud University-Computer and
  Information Sciences}, vol.~32, no.~1, pp. 99--112, 2020.

\bibitem{tan2016phishwho}
C.~L. Tan, K.~L. Chiew, K.~Wong \emph{et~al.}, ``Phishwho: Phishing webpage
  detection via identity keywords extraction and target domain name finder,''
  \emph{Decision Support Systems}, vol.~88, pp. 18--27, 2016.

\bibitem{shirazi2018kn0w}
H.~Shirazi, B.~Bezawada, and I.~Ray, ``" kn0w thy doma1n name" unbiased
  phishing detection using domain name based features,'' in \emph{Proceedings
  of the 23nd ACM on Symposium on Access Control Models and Technologies},
  2018, pp. 69--75.

\bibitem{nguyen2013detecting}
L.~A.~T. Nguyen, B.~L. To, H.~K. Nguyen, and M.~H. Nguyen, ``Detecting phishing
  web sites: A heuristic url-based approach,'' in \emph{2013 International
  Conference on Advanced Technologies for Communications (ATC 2013)}.\hskip 1em
  plus 0.5em minus 0.4em\relax IEEE, 2013, pp. 597--602.

\bibitem{babagoli2019heuristic}
M.~Babagoli, M.~P. Aghababa, and V.~Solouk, ``Heuristic nonlinear regression
  strategy for detecting phishing websites,'' \emph{Soft Computing}, vol.~23,
  no.~12, pp. 4315--4327, 2019.

\bibitem{jeeva2016intelligent}
S.~C. Jeeva and E.~B. Rajsingh, ``Intelligent phishing url detection using
  association rule mining,'' \emph{Human-centric Computing and Information
  Sciences}, vol.~6, no.~1, pp. 1--19, 2016.

\bibitem{sreedharan2016systems}
J.~Sreedharan and R.~Mohandas, ``Systems and methods for risk rating and
  pro-actively detecting malicious online ads,'' apr 5 2016, uS Patent
  9,306,968.

\bibitem{liang2016cracking}
B.~Liang, M.~Su, W.~You, W.~Shi, and G.~Yang, ``Cracking classifiers for
  evasion: a case study on the google's phishing pages filter,'' in
  \emph{Proceedings of the 25th International Conference on World Wide Web},
  2016, pp. 345--356.

\bibitem{rajivan2018creative}
P.~Rajivan and C.~Gonzalez, ``Creative persuasion: a study on adversarial
  behaviors and strategies in phishing attacks,'' \emph{Frontiers in
  psychology}, vol.~9, p. 135, 2018.

\bibitem{royremains}
S.~S. Roy, U.~Karanjit, and S.~Nilizadeh, ``What remains uncaught?:
  Characterizing sparsely detected malicious urls on twitter.''

\bibitem{aleroud2020bypassing}
A.~AlEroud and G.~Karabatis, ``Bypassing detection of url-based phishing
  attacks using generative adversarial deep neural networks,'' in
  \emph{Proceedings of the Sixth International Workshop on Security and Privacy
  Analytics}, 2020, pp. 53--60.

\bibitem{oest2020phishtime}
A.~Oest, Y.~Safaei, P.~Zhang, B.~Wardman, K.~Tyers, Y.~Shoshitaishvili, and
  A.~Doup{\'e}, ``Phishtime: Continuous longitudinal measurement of the
  effectiveness of anti-phishing blacklists,'' in \emph{29th $\{$USENIX$\}$
  Security Symposium ($\{$USENIX$\}$ Security 20)}, 2020, pp. 379--396.

\bibitem{twitter:2021}
Twitter, \url{https://twitter.com/home}.

\bibitem{14phishing}
``{1.4 million phishing websites are created every month},''
  \url{https://zd.net/30T55DW}, 2020.

\bibitem{mohammad2014intelligent}
R.~M. Mohammad, F.~Thabtah, and L.~McCluskey, ``Intelligent rule-based phishing
  websites classification,'' \emph{IET Information Security}, vol.~8, no.~3,
  pp. 153--160, 2014.

\bibitem{vayansky2018phishing}
I.~Vayansky and S.~Kumar, ``Phishing--challenges and solutions,''
  \emph{Computer Fraud \& Security}, vol. 2018, no.~1, pp. 15--20, 2018.

\bibitem{junger2017priming}
M.~Junger, L.~Montoya, and F.-J. Overink, ``Priming and warnings are not
  effective to prevent social engineering attacks,'' \emph{Computers in human
  behavior}, vol.~66, pp. 75--87, 2017.

\bibitem{bullee2018anatomy}
J.-W.~H. Bull{\'e}e, L.~Montoya, W.~Pieters, M.~Junger, and P.~Hartel, ``On the
  anatomy of social engineering attacks—a literature-based dissection of
  successful attacks,'' \emph{Journal of investigative psychology and offender
  profiling}, vol.~15, no.~1, pp. 20--45, 2018.

\bibitem{alsharnouby2015phishing}
M.~Alsharnouby, F.~Alaca, and S.~Chiasson, ``Why phishing still works: User
  strategies for combating phishing attacks,'' \emph{International Journal of
  Human-Computer Studies}, vol.~82, pp. 69--82, 2015.

\bibitem{arachchilage2014security}
N.~A.~G. Arachchilage and S.~Love, ``Security awareness of computer users: A
  phishing threat avoidance perspective,'' \emph{Computers in Human Behavior},
  vol.~38, pp. 304--312, 2014.

\bibitem{abroshan2021phishing}
H.~Abroshan, J.~Devos, G.~Poels, and E.~Laermans, ``Phishing happens beyond
  technology: The effects of human behaviors and demographics on each step of a
  phishing process,'' \emph{IEEE Access}, vol.~9, pp. 44\,928--44\,949, 2021.

\bibitem{phishingfinance:2020}
C.~Greg~Iacurci, ``Coronavirus scams, feeding off investor fears, mimic fraud
  from the 2008 financial crisis,''
  \url{https://www.cnbc.com/2020/03/20/coronavirus-scams-on-the-rise-mimic-fraud-in-2008-financial-crisis.html}.

\bibitem{ahmad2020corona}
T.~Ahmad, ``Corona virus (covid-19) pandemic and work from home: Challenges of
  cybercrimes and cybersecurity,'' \emph{Available at SSRN 3568830}, 2020.

\bibitem{covidphishincrease:2020}
P.~Jason~Cohen, ``Phishing attacks increase 350 percent amid covid-19
  quarantine,''
  \url{https://www.pcmag.com/news/phishing-attacks-increase-350-percent-amid-covid-19-quarantine}.

\bibitem{phishtank}
``{PhishTank},'' \url{https://www.phishtank.com/faq.php}, 2020.

\bibitem{openphish}
Openphish, ``Phishing feed,'' \url{"https://openphish.com/faq.html" }.

\bibitem{sheng2009empirical}
S.~Sheng, B.~Wardman, G.~Warner, L.~Cranor, J.~Hong, and C.~Zhang, ``An
  empirical analysis of phishing blacklists,'' 2009.

\bibitem{bell2020analysis}
S.~Bell and P.~Komisarczuk, ``An analysis of phishing blacklists: Google safe
  browsing, openphish, and phishtank,'' in \emph{Proceedings of the
  Australasian Computer Science Week Multiconference}, 2020, pp. 1--11.

\bibitem{moore2008evaluating}
T.~Moore and R.~Clayton, ``Evaluating the wisdom of crowds in assessing
  phishing websites,'' in \emph{International Conference on Financial
  Cryptography and Data Security}.\hskip 1em plus 0.5em minus 0.4em\relax
  Springer, 2008, pp. 16--30.

\bibitem{firefox:2021}
M.~Firefox, ``How does built-in phishing and malware protection work?'' \url{
  https://support.mozilla.org/en-US/kb/how-does-phishing-and-malware-protection-work}.

\bibitem{operafeed:2020}
Opera, ``Opera introduces fraud protection, powered by geotrust and phishtank:
  New release expands opera’s commitment to secure browsing,''
  \url{https://bit.ly/3FkKBTx}.

\bibitem{safarifeed:2020}
Apple, ``Safari and privacy,'' \url{
  https://www.apple.com/legal/privacy/data/en/safari/ }.

\bibitem{phishfriends}
``{Friends of PhishTank [Infographic]},''
  \url{https://www.phishtank.com/friends.php}, 2020.

\bibitem{Twitter:2020}
Twitter, \url{https://developer.twitter.com/en}.

\bibitem{urldefanging}
``{Email Security – Defanging URLs},''
  \url{https://www.ibm.com/docs/en/rsoa-and-rp/32.0?topic=SSBRUQ_32.0.0/com.ibm.resilient.doc/install/resilient_install_defangURLs.htm},
  2021.

\bibitem{VirusTotal:2020}
VirusTotal, \url{https://www.virustotal.com/gui/home/upload}.

\bibitem{virustotal}
``{VirusTotal},'' \url{https://www.virustotal.com/gui/home/}, 2020.

\bibitem{masri2017automated}
R.~Masri and M.~Aldwairi, ``Automated malicious advertisement detection using
  virustotal, urlvoid, and trendmicro,'' in \emph{2017 8th International
  Conference on Information and Communication Systems (ICICS)}.\hskip 1em plus
  0.5em minus 0.4em\relax IEEE, 2017, pp. 336--341.

\bibitem{tanaka2017analysis}
Y.~Tanaka, M.~Akiyama, and A.~Goto, ``Analysis of malware download sites by
  focusing on time series variation of malware,'' \emph{Journal of
  computational science}, vol.~22, pp. 301--313, 2017.

\bibitem{sun2016automating}
B.~Sun, M.~Akiyama, T.~Yagi, M.~Hatada, and T.~Mori, ``Automating url blacklist
  generation with similarity search approach,'' \emph{IEICE TRANSACTIONS on
  Information and Systems}, vol.~99, no.~4, pp. 873--882, 2016.

\bibitem{wang2019rmvdroid}
H.~Wang, J.~Si, H.~Li, and Y.~Guo, ``Rmvdroid: towards a reliable android
  malware dataset with app metadata,'' in \emph{2019 IEEE/ACM 16th
  International Conference on Mining Software Repositories (MSR)}.\hskip 1em
  plus 0.5em minus 0.4em\relax IEEE, 2019, pp. 404--408.

\bibitem{whois}
R.~Penman, ``Python-whois,'' \url{https://pypi.org/project/python-whois/}.

\bibitem{lehmann2012dynamical}
J.~Lehmann, B.~Gon{\c{c}}alves, J.~J. Ramasco, and C.~Cattuto, ``Dynamical
  classes of collective attention in twitter,'' in \emph{Proceedings of the
  21st international conference on World Wide Web}, 2012, pp. 251--260.

\bibitem{ferreira2015analysis}
A.~Ferreira and G.~Lenzini, ``An analysis of social engineering principles in
  effective phishing,'' in \emph{2015 Workshop on Socio-Technical Aspects in
  Security and Trust}.\hskip 1em plus 0.5em minus 0.4em\relax IEEE, 2015, pp.
  9--16.

\bibitem{provos2008all}
N.~Provos, P.~Mavrommatis, M.~Rajab, and F.~Monrose, ``All your iframes point
  to us,'' 2008.

\bibitem{peng2019opening}
P.~Peng, L.~Yang, L.~Song, and G.~Wang, ``Opening the blackbox of virustotal:
  Analyzing online phishing scan engines,'' in \emph{Proceedings of the
  Internet Measurement Conference}, 2019, pp. 478--485.

\bibitem{ciscotrend}
``{2021 Cybersecurity threat trends: phishing, crypto top the list},''
  \url{https://umbrella.cisco.com/info/2021-cyber-security-threat-trends-phishing-crypto-top-the-list},
  2021.

\bibitem{trojanhorses}
``{What is a Trojan? Is it a virus or is it malware?}''
  \url{https://us.norton.com/internetsecurity-malware-what-is-a-trojan.html},
  2020.

\bibitem{androidmalware}
``{Mobile Malware},''
  \url{https://usa.kaspersky.com/resource-center/threats/mobile}, 2020.

\bibitem{cryptojacking}
``{Cryptojacking – What is it?}''
  \url{https://www.malwarebytes.com/cryptojacking}, 2020.

\bibitem{browserhijacker}
``{What are Browser Hijackers?}''
  \url{https://us.norton.com/internetsecurity-malware-what-are-browser-hijackers.html},
  2020.

\bibitem{keylogger}
``{Demystifying a Keylogger – How They Monitor What You Type and What You Can
  Do About It?}''
  \url{https://home.sophos.com/en-us/security-news/2019/what-is-a-keylogger.aspx},
  2020.

\bibitem{scareware}
``{What is Scareware?}''
  \url{https://usa.kaspersky.com/resource-center/definitions/scareware}, 2020.

\bibitem{ransomware}
``{What is Ransomware?}''
  \url{https://www.mcafee.com/enterprise/en-us/security-awareness/ransomware.html},
  2020.

\bibitem{vtpaper_blackbox}
P.~Peng, L.~Yang, L.~Song, and G.~Wang, ``Opening the blackbox of virustotal:
  Analyzing online phishing scan engines,'' in \emph{Proceedings of the
  Internet Measurement Conference}, 2019, pp. 478--485.

\bibitem{npapernot}
N.~Papernot, ``Detecting phishing websites using a decision tree,''
  \url{"https://github.com/npapernot/phishing-detection"}.

\bibitem{mustafadataset}
R.~M.~A. Mohammad, ``Phishing websites data set,''
  \url{https://openphish.com/faq.html}.

\bibitem{sharkcop}
C.~H. Tung, ``Sharkcop,'' \url{https://github.com/CaoHoangTung/sharkcop}.

\bibitem{sharkcoppress}
T.~D. Swig, ``Sharkcop,'' \url{https://bit.ly/3spN1wz" }.

\bibitem{mcshane2021emoji}
L.~McShane, E.~Pancer, M.~Poole, and Q.~Deng, ``Emoji, playfulness, and brand
  engagement on twitter,'' \emph{Journal of Interactive Marketing}, vol.~53,
  pp. 96--110, 2021.

\bibitem{kang2012modeling}
B.~Kang, J.~O'Donovan, and T.~H{\"o}llerer, ``Modeling topic specific
  credibility on twitter,'' in \emph{Proceedings of the 2012 ACM international
  conference on Intelligent User Interfaces}, 2012, pp. 179--188.

\bibitem{gupta2018framework}
H.~Gupta, M.~S. Jamal, S.~Madisetty, and M.~S. Desarkar, ``A framework for
  real-time spam detection in twitter,'' in \emph{2018 10th International
  Conference on Communication Systems \& Networks (COMSNETS)}.\hskip 1em plus
  0.5em minus 0.4em\relax IEEE, 2018, pp. 380--383.

\bibitem{herzallah2018feature}
W.~Herzallah, H.~Faris, and O.~Adwan, ``Feature engineering for detecting
  spammers on twitter: Modelling and analysis,'' \emph{Journal of Information
  Science}, vol.~44, no.~2, pp. 230--247, 2018.

\bibitem{mcknight2010mann}
P.~E. McKnight and J.~Najab, ``Mann-whitney u test,'' \emph{The Corsini
  encyclopedia of psychology}, pp. 1--1, 2010.

\end{thebibliography}
